%% file: manuscript.tex
\pdfoutput=1
\documentclass[sigconf]{acmart}

\usepackage[english]{babel}

\copyrightyear{2026}
\acmYear{2026}
\setcopyright{cc}
\setcctype{by}

\acmConference[SoCC '26]{ACM Symposium on Cloud Computing V.2}{November 18--20, 2026}{Singapore, Singapore}
\acmBooktitle{ACM Symposium on Cloud Computing V.2 (SoCC '26), November 18--20, 2026, Singapore, Singapore}
\acmDOI{10.1145/3828161.3857852}
\acmISBN{979-8-4007-2816-7/2026/11}

\input{./sections/package.tex}

\newcommand{\sys}{NemoANN\xspace}
\newcommand{\pipeann}{PipeANN\xspace}
\newcommand{\pipeannfilter}{PipeANN-Filter\xspace}
\newcommand{\diskann}{DiskANN\xspace}
\newcommand{\fdiskann}{F-DiskANN\xspace}

\begin{document}
\title{\sys: I/O-Efficient Filtered Vector Search on SSDs with Graph Tunneling}
\author{Nakyung Lee}
\authornote{Nakyung Lee and Soobin Cho contributed equally to this work.}
\affiliation{%
  \institution{Sungshin Women's University}
  \country{Republic of Korea}
}
\email{220254009@sungshin.ac.kr}

\author{Soobin Cho}
\authornotemark[1]
\affiliation{%
  \institution{Sungshin Women's University}
  \country{Republic of Korea}
}
\email{20231163@sungshin.ac.kr}

\author{Jiwoong Park}
\affiliation{%
  \institution{Sungshin Women's University}
  \country{Republic of Korea}
}
\email{jwpark12@sungshin.ac.kr}

\author{Gyuyeong Kim}
\affiliation{%
  \institution{Yonsei University}
  \country{Republic of Korea}
}
\email{gyuyeong.kim@yonsei.ac.kr}

\renewcommand{\shortauthors}{Lee et al.}
\begin{abstract}
\input{./sections/abstract.tex}
\end{abstract}

\maketitle

\input{./sections/introduction.tex}
\input{./sections/background.tex}

\input{./sections/design.tex}

\input{./sections/implementation.tex}

\input{./sections/evaluation.tex}
\input{./sections/related.tex}
\input{./sections/conclusion.tex}
\label{lastpage1}
\bibliographystyle{ACM-Reference-Format}
\bibliography{./nemoann}
\label{lastpage2}

\end{document}

%% file: sections/package.tex
\usepackage{tikz}
\usepackage{xcolor,xspace,booktabs}
\PassOptionsToPackage{hyphens}{url}
\usepackage{graphicx,subfig,multirow,amsmath,algorithm,algpseudocode}
\usepackage{pifont}
\usepackage{textcomp}
\usepackage{enumitem}
\usepackage{color,soul}
\usepackage{amsthm}

%% file: sections/abstract.tex
We introduce NemoANN, an I/O-efficient SSD-based Approximate Nearest Neighbor Search (ANNS) system for filtered search.
Existing SSD-based systems either waste I/O on post-filtering or require expensive filter-aware index rebuilds.
Our key insight is that a non-matching node can never be a result, so the search needs only its small neighbor list, not the full-precision vector on SSD.
NemoANN therefore checks each predicate before the SSD read and tunnels through non-matching nodes using in-memory routing metadata, preserving graph connectivity without modifying the index.
On a 100M-vector benchmark, NemoANN reduces SSD reads by about 10$\times$ and improves throughput by 8.0$\times$ over post-filtering and by 5.0$\times$ over PipeANN-Filter, a state-of-the-art filtered SSD system, at 10\% selectivity and 90\% recall.
The gains over post-filtering grow at lower selectivity, reaching 13.1$\times$ at 5\% selectivity.

%% file: sections/introduction.tex
\section{Introduction}\label{sec:intro}
High-dimensional vectors have become a universal representation for unstructured data across diverse applications, from recommendation systems to retrieval-augmented generation (RAG)~\cite{hnsw,cacheblend,hedrarag,metis-rag}.
At the billion-scale, exact search is prohibitively expensive, so approximate nearest neighbor search (ANNS) with graph-based indexes is widely adopted for its superior latency-recall tradeoff~\cite{hnsw,pathweaver,cxlanns,quake}.
Recently, to scale beyond memory capacity, SSD-based ANNS systems (e.g., \diskann~\cite{diskann}, \pipeann~\cite{pipeann}) have been proposed, making large-scale vector search practical on a single commodity server.
These systems store full-precision vectors and graph topology on SSDs, retaining only compressed vectors in memory for approximate distance computation.

In practice, vector search queries rarely target the entire dataset.
Instead, they commonly restrict results using metadata predicates such as document type, time range, or product category~\cite{milvus,pinecone}.
General-purpose SSD-based ANNS systems such as \diskann and \pipeann handle such \emph{filtered search} through post-filtering. 
They fetch candidate nodes from SSD, compute exact distances, and then discard nodes that do not satisfy the predicate.
This approach preserves the original graph traversal but wastes I/O and computation on non-matching nodes.
For example, at 10\% selectivity, up to 90\% of SSD reads and exact-distance computations may be spent on vectors that cannot appear in the final result.

Filtering before SSD access could eliminate this waste, but directly skipping non-matching nodes (na\"ive joint-filtering) breaks graph traversal~\cite{filtering-ann-study,vecbench}.
A skipped node cannot be expanded, even though its neighbors may provide essential paths to matching results.
Filtered-ANNS systems restore this connectivity by changing the index itself.
Filter-aware graphs such as \fdiskann~\cite{filtered-diskann} build label-specific edges and entry points, so they avoid reads of non-matching nodes, but they support only the label space fixed at build time and must be rebuilt when the filter schema changes.
Predicate-agnostic designs instead densify the graph with additional edges so that the search can skip most non-matching nodes without losing connectivity~\cite{acorn,pipeann-filter}.
On SSD, \pipeannfilter~\cite{pipeann-filter} adds 2-hop neighbors to each node record and speculatively expands nodes that pass in-memory probabilistic filters.
This keeps predicates flexible with little memory, but enlarges the on-disk index and still reads non-matching nodes, both false positives and bridges, from SSD.
Existing systems therefore pay for filtering in one of three places: wasted SSD reads (post-filtering), predicate flexibility and rebuild cost (filter-aware graphs), or a larger index with speculative reads (densified graphs).
In this context, we ask: \emph{Can filtered vector search avoid unnecessary SSD I/O on an unmodified graph index and on-disk layout while leaving the predicate to query time?}

To answer this question, we present \sys, an I/O-efficient SSD-based graph ANNS system for filtered search.
Our key insight is that the search needs a non-matching node's neighbors, not its vector.
Such a node cannot appear in the results, so the search visits it only as a waypoint to its neighbors.
The visit needs only the node’s neighbor list and an approximate distance, and both fit in memory.
Existing SSD-based systems co-locate a node's vector and neighbor list so that a single aligned read supports both graph traversal and exact-distance computation; as a result, even a waypoint visit costs a full SSD read.

\sys therefore keeps a compact copy of the graph's neighbor lists in memory as the neighbor store.
SSD-based ANNS systems already keep the compressed vectors in memory, so the neighbor store completes the in-memory information for traversing a non-matching node.
During traversal, \sys checks the filter before issuing an SSD read.
A matching node is fetched for exact-distance computation.
A non-matching node is not fetched; \sys moves the search through it using its in-memory neighbor list and approximate distance.
We call this \emph{graph tunneling}.
Unlike densified graphs, tunneling routes over the original edges and leaves the index and its on-disk layout unchanged.
Unlike speculative filtering, it checks the exact predicate before I/O, so non-matching nodes still bridge the search but never cost an SSD read.
\sys thus trades memory for SSD I/O; for 100M vectors, the neighbor store requires 12.3\,GB at the default setting and can be reduced to about 3.4\,GB while still outperforming post-filtering baselines.

We evaluate \sys on datasets ranging from 10M to 1B vectors with both synthetic and real multi-label metadata~\cite{bigann-neurips23}.
On BigANN-100M, \sys reduces SSD I/Os by about 10$\times$ relative to post-filter baselines and improves throughput by 8.0$\times$ at 10\% selectivity and 90\% recall.
As filters become more selective, the gains grow further, reaching 13.1$\times$ at 5\% selectivity.
Against \pipeannfilter, \sys achieves 5.0$\times$ higher throughput at 10\% selectivity while keeping the original index, which is 14$\times$ smaller on SSD than \pipeannfilter's densified index.
\sys also closes the gap with in-memory search; it matches or surpasses Vamana in single-thread latency despite using an SSD and 3.4$\times$ less memory.

In summary, this paper makes the following contributions:
\begin{itemize}[noitemsep,leftmargin=*]
\item We characterize the I/O dilemma in filtered SSD-based graph search, where post-filtering wastes SSD I/O and computation while na\"ive joint-filtering breaks graph connectivity and severely degrades recall (\S\ref{sec:bgm}).

\item We present \sys, which turns a pre-I/O filter failure into a tunneling decision that serves the node from its in-memory neighbor list, enabling I/O-efficient filtered search on an unmodified graph index (\S\ref{sec:design}).

\item We evaluate \sys on datasets from 10M to 1B vectors with synthetic and real metadata, showing about 10$\times$ fewer SSD I/Os and 8.0$\times$ higher throughput than SSD-based post-filter baselines at 10\% selectivity and matched recall (13.1$\times$ at 5\% selectivity), and 5.0$\times$ higher throughput than \pipeannfilter (\S\ref{sec:eval}).
\end{itemize}

%% file: sections/background.tex
\section{Background and Motivation}\label{sec:bgm}
We begin by reviewing SSD-based graph ANNS (\S\ref{sec:bg-ssd}), then explain the I/O dilemma in filtered vector search (\S\ref{sec:io-waste}). We next motivate the need for I/O-efficient joint-filtering on unmodified graph indexes (\S\ref{sec:toward}).

\subsection{SSD-Based Graph ANNS}\label{sec:bg-ssd}
Graph-based approximate nearest neighbor search (ANNS) organizes vectors as a proximity graph, where each vector is a node and edges connect approximate neighbors.
Given a query, the search starts from an entry point and greedily traverses toward nodes that appear progressively closer to the query, returning the closest visited nodes as approximate nearest neighbors~\cite{hnsw,diskann,pipeann}.

At the billion scale, storing full-precision vectors in memory is often infeasible on a commodity server~\cite{cxlanns,smartanns,fusionanns,spfresh,pimann}.
For example, storing 128-dimensional float32 vectors together with a degree-64 graph at a scale of one billion points requires approximately 768\,GB of memory, far exceeding the memory capacity of a commodity server.
This motivates SSD-based graph ANNS systems such as \diskann~\cite{diskann} that place node records on NVMe SSDs while keeping only compressed vectors in memory.
Product Quantization (PQ) encodes each vector as a short code (e.g., 32 bytes), enabling approximate distance computation without an SSD read.
During search, these PQ distances prioritize which candidates to expand, while exact distances from the full-precision vectors on SSD determine the final results.
In \diskann, each node occupies a 4\,KB-aligned record containing its raw vector and neighbor list.
Search proceeds in rounds. The system selects a beam of promising candidates, issues SSD reads for them, waits for the batch to finish, and then continues.

\pipeann~\cite{pipeann} replaces this batch-and-wait design by pipelining reads with \texttt{io\_uring}.
It continuously submits reads, polls completions, and processes nodes with many I/Os in flight.
This overlap substantially improves throughput.
However, in both systems, graph traversal still incurs a random SSD read for every expanded node, and each read costs tens to hundreds of microseconds.
As a result, per-node SSD I/O remains the dominant cost.

\begin{figure}[t]
\centering
\subfloat[Throughput scaling]{\includegraphics[width=0.49\linewidth]{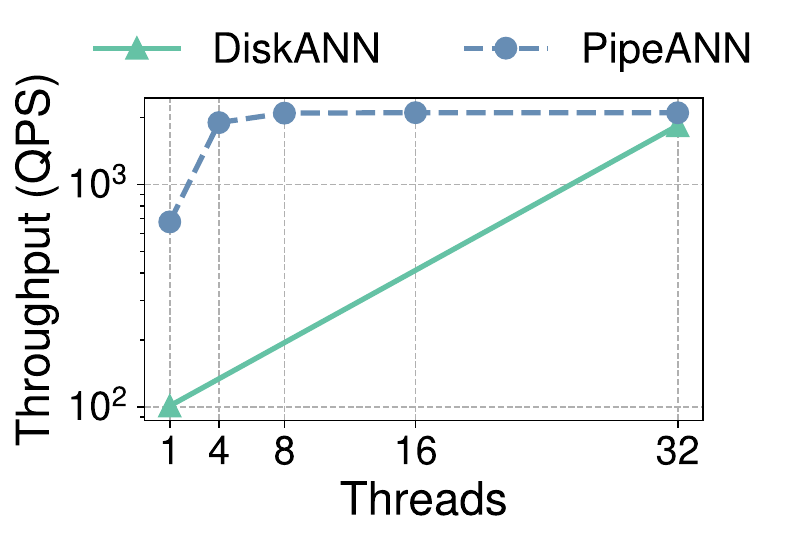}}\hfill
\subfloat[Post-filter vs.\ na\"ive joint-filter]{\includegraphics[width=0.49\linewidth]{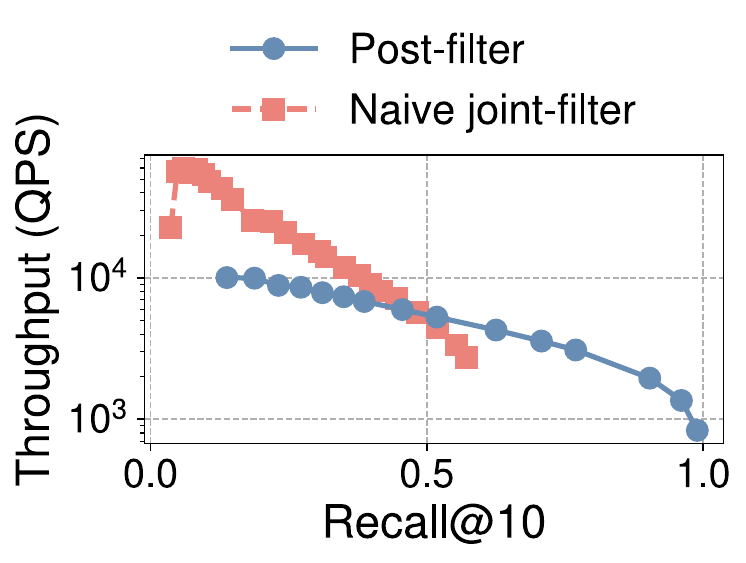}}
\caption{Motivating experiments on BigANN-100M with 10\% selectivity. (a)~Post-filtering systems plateau early because of wasted per-node work. (b)~Na\"ive joint-filtering destroys graph connectivity, degrading throughput and recall.}
\label{fig:motivation}
\end{figure}

\subsection{The I/O Dilemma in Filtered Search}\label{sec:io-waste}

\noindent\textbf{Filtered vector search is common.}
Production queries rarely search the entire dataset.
Enterprise RAG restricts by document type and access control, legal search filters by jurisdiction and time range, and e-commerce search narrows by product category.
Such metadata predicates are pervasive in practice~\cite{milvus,pinecone,vbase,acorn}.
The \emph{selectivity} $s$ of a filter is the fraction of the dataset that satisfies the filter.
Selectivities in the 1--10\% range are common in both benchmark and production settings~\cite{curator,azure-ai-search,vdbbench}.

\noindent\textbf{Post-filtering wastes resources.}
Existing SSD-based graph ANNS systems such as \diskann and \pipeann handle filters by \emph{post-filtering}. 
They read every node selected for expansion from SSD and only then check the predicate.
This approach is general, but it pays full per-node cost (SSD I/O, exact distance computation, and candidate management) regardless of whether the node matches the filter.
This waste becomes costly under multi-threads.
In Figure~\ref{fig:motivation}~(a), \pipeann initially outperforms \diskann by overlapping I/O and computation.
However, as concurrency increases, they converge because both ultimately spend most of their budget processing nodes that fail the filter.

\noindent\textbf{Na\"ive joint-filtering breaks graph search.}
A natural alternative is to push the predicate before I/O and skip non-matching nodes during traversal, which is called \emph{joint-filtering} (also known as in-filtering~\cite{pipeann-filter}).
However, this is fundamentally problematic for graph search, because non-matching nodes are not only poor result candidates but also routing waypoints that connect the search to other regions of the graph.
Skipping their expansion prevents the search from reaching their neighbors, even when those neighbors match the predicate.
Because many paths between matching nodes pass through non-matching nodes, na\"ive joint-filtering on an unmodified graph fragments the set of matching nodes into many small disconnected components, particularly at low selectivity.

Figure~\ref{fig:motivation}(b) confirms this.
At the same throughput, na\"ive joint-filtering achieves much lower recall than post-filtering, and its maximum recall plateaus at ${\sim}$57\% even at 2.7K QPS, compared to $>$99\% for post-filtering.
Increasing the search list size $L$ cannot overcome this limitation since unreachable components remain invisible regardless of how many candidates the traversal explores.

\noindent\textbf{Filter-aware indexes sacrifice flexibility.}\label{sec:rebuild-dilemma}
Prior work on filtered search incorporates label information into the index structure.
For example, Filtered-DiskANN (\fdiskann)~\cite{filtered-diskann} augments the graph with label-aware connectivity and entry points, eliminating I/O for non-matching nodes.
However, they require the label space to be known at build time.
Supporting a new predicate or changing the taxonomy requires rebuilding the index, which can take days at the billion scale.
They also cannot handle ad-hoc predicates such as subset predicates over multi-label metadata or range predicates over continuous attributes.
In-memory systems follow the same direction by fusing attributes into the graph or by maintaining filter-specific subindexes~\cite{nhq,ung,curator,sieve}.

\noindent\textbf{Densified graphs trade index size for connectivity.}
Predicate-agnostic designs keep a single index for all predicates and instead add edges so that the search can skip most non-matching nodes without losing connectivity.
ACORN~\cite{acorn} densifies an in-memory graph for this purpose.
PipeANN-Filter~\cite{pipeann-filter} appends a random sample of 2-hop neighbors to each on-disk record, identifies a superset of matching nodes with in-memory probabilistic filters such as per-node Bloom filters, and verifies the exact attributes stored in each record when results are reranked.
A per-query cost model chooses among this speculative in-filtering, pre-filtering, and post-filtering.
This design keeps predicates flexible and its in-memory filter state small, but it pays in two ways.
First, each densified record is larger than the original one, which enlarges both the on-disk index and every read.
Second, the search still reads non-matching nodes from SSD, namely false positives of the probabilistic filters and the non-matching neighbors it deliberately explores as bridges when too few matching neighbors are found.

\begin{figure}[t!]
\centering\hfill
\includegraphics[width=0.95\linewidth]{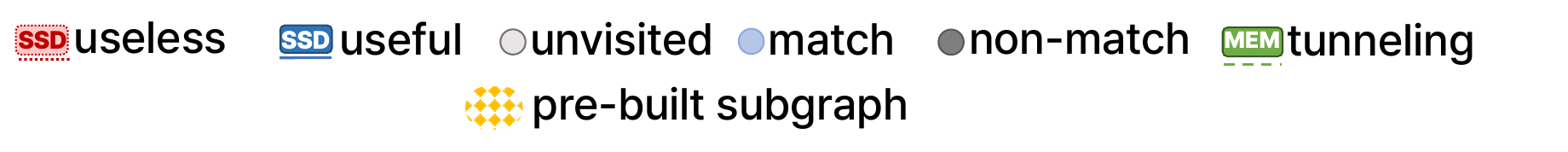}\\[-2pt]
\subfloat[Post-filtering]{\includegraphics[width=0.3\linewidth]{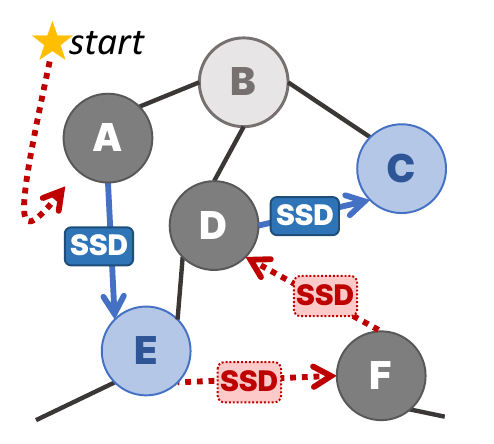}}\hfill
\subfloat[Filter-aware index]{\includegraphics[width=0.3\linewidth]{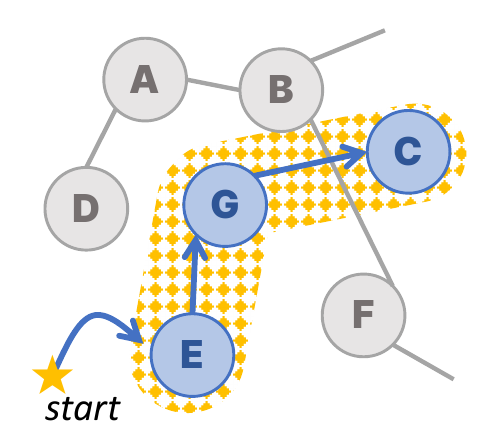}}\hfill
\subfloat[\sys]{\includegraphics[width=0.3\linewidth]{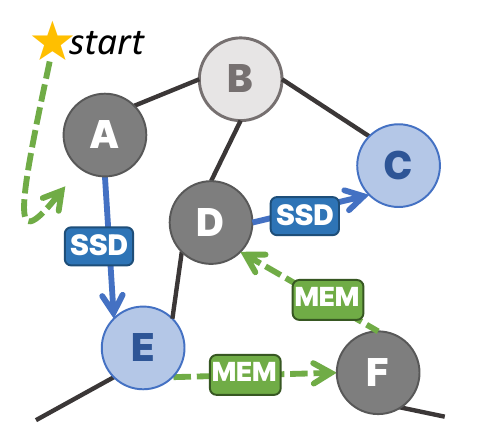}}\hfill
\caption{Different approaches for filtered search.
\label{fig:motiv-com}}
\end{figure}

\subsection{Toward I/O-Efficient Joint-filtering}\label{sec:toward}
\noindent\textbf{Rethinking why joint-filtering seems impractical.}
Checking the filter before I/O is not a new goal in itself; \fdiskann avoids reads for non-matching nodes by rebuilding the graph around a fixed label space, and \pipeannfilter avoids most of them by densifying every records.
On an unmodified graph and on-disk layout, however, skipping a non-matching node discards a useful waypoint, so a system must either read the node anyway or accept the loss of connectivity.

The problem is not joint-filtering itself, but the coupling between traversal and I/O in existing systems.
Specifically, a node's neighbor list is stored alongside its vector in a single SSD record; unless that record is cached in memory, discovering the node's neighbors requires reading it from the SSD.
This means that even determining the next traversal hop requires a disk I/O, regardless of whether the node matches the filter.

\noindent\textbf{Insight: the missing piece is small.}
Graph traversal requires only two pieces of per-node information, an approximate distance and a neighbor list.
Existing SSD-based systems already keep compressed vectors in memory to estimate distances, but store each neighbor list on SSD together with the full-precision vector.
Thus, to traverse a node without an SSD read, the only missing information is the neighbor list.
Because neighbor lists are much smaller than full-precision vectors, a compact copy of the graph adjacency can be kept in memory at a practical cost.

\noindent\textbf{Our approach.}
Based on this observation, \sys maintains a compact in-memory copy of each node's neighbor list.
Before issuing an SSD read, \sys evaluates the filter predicate using in-memory metadata.
If the node matches the predicate, \sys fetches its full-precision vector from SSD and computes its exact distance.
Otherwise, \sys skips the SSD read but continues traversal using the node's in-memory compressed vector and neighbor list.
We call this routing through non-matching nodes without SSD access \emph{graph tunneling}.
Unlike speculative filtering, the check is exact, so a non-matching node never costs an SSD read, whether or not the search routes through it.
Because \sys uses the original graph without rebuilding or modifying it for specific filters, it supports arbitrary predicates known only at query time.

\begin{table}[t]
\centering
\setlength{\tabcolsep}{3pt}
\caption{Comparison with existing works.\label{tab:comparison}}
\footnotesize
\begin{tabular}{@{}lcccc@{}}
\hline\hline
 &
\shortstack{\pipeannfilter~\cite{pipeann-filter}} &
\shortstack{\fdiskann~\cite{filtered-diskann}} &
\shortstack{\textbf{\sys}} \\ \hline\hline
Filter type & Speculative-filtering     &  Filter-aware Index  & Joint-filtering \\
I/O-efficient    & $\times$ &  $\surd$ & $\surd$ \\
Any filter predicates  & $\surd$  &  $\times$ & $\surd$ \\
No index rebuild & $\surd$ &  $\times$ & $\surd$ \\
\hline
\end{tabular}
\vspace{-8pt}
\end{table}

\noindent\textbf{Comparison to existing works.}
Figure~\ref{fig:motiv-com} and Table~\ref{tab:comparison} summarize how \sys differs from prior work.
Post-filtering~\cite{pipeann,diskann} reads every traversed node from SSD and applies the predicate only afterward, while a filter-aware index~\cite{filtered-diskann} eliminates I/O for non-matching nodes but fixes the label space at build time and requires a rebuild for new predicates.
PipeANN-Filter~\cite{pipeann-filter} keeps predicates flexible with small in-memory state, but enlarges every on-disk record and still reads false positives and bridge nodes.
\sys checks the exact predicate before I/O and traverses non-matching nodes in memory, preserving connectivity and eliminating their SSD reads without restricting the predicate or changing the index; the cost is the DRAM for the neighbor and filter stores.

%% file: sections/design.tex
\begin{figure*}[t]
\centering
\includegraphics[width=0.8\linewidth]{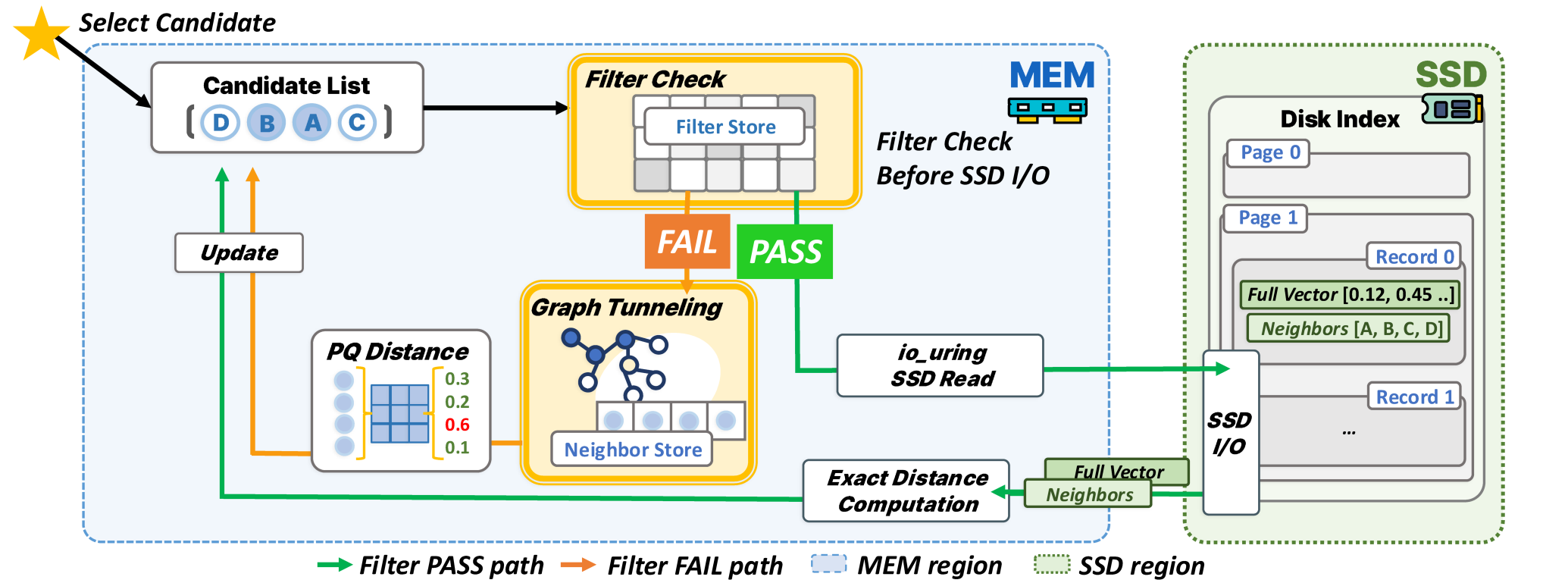}
\caption{\sys overview.
A candidate is first checked against the in-memory filter store.
Filter-passing nodes follow the normal SSD path, an asynchronous read followed by exact distance computation.
Filter-failing nodes take the graph tunneling path. The neighbor store provides the adjacency list, PQ distances are computed in memory, and promising neighbors are inserted back into the candidate list.
Both paths feed into the same sorted frontier.
}
\label{fig:overview}
\end{figure*}

\section{\sys Design}\label{sec:design}

\subsection{Overview}\label{sec:overview-arch}
Figure~\ref{fig:overview} shows the architecture of \sys.
The central design principle is to serve a node's two roles separately.
Whether a node's full-precision vector is fetched from SSD should depend on whether the node can contribute to the query result, not on whether the node lies on the search path.
A node that fails the filter retains only its routing role, and \sys serves that role entirely from memory.

\sys realizes this by inserting a filter-aware dispatch stage into the asynchronous search pipeline.
Each candidate is dispatched to one of two paths based on a lightweight in-memory predicate check.
The two paths differ only in how a node is processed.
Both feed candidates into the same sorted frontier, so the rest of the search algorithm is unaffected.
This preserves the asynchronous I/O pipeline~\cite{pipeann} intact while ensuring that every SSD read serves a node that can appear in the final result.
The following subsections describe each component in detail.

\subsection{Key Components}\label{sec:mechanism}
\sys relies on two in-memory structures.
The \emph{filter store} decides which path a candidate takes, and the \emph{neighbor store} makes the second path possible.

\noindent\textbf{Filter store.}\label{sec:preio}
The \emph{filter store} is a memory-resident structure that holds each node's filter metadata and evaluates the predicate after an $O(1)$ lookup by node ID.
The evaluation is a direct check on the node's stored metadata rather than a probabilistic test such as a Bloom filter; exactness matters because an erroneous pass would waste an SSD read and an erroneous fail would silently exclude a valid result.
An equality filter compares a single fixed-width label in constant time, while a multi-label subset filter tests each query tag against the node's in-memory tag set, so its cost grows with the number of query tags.
Even this most expensive case is a short in-memory loop, far cheaper than the SSD read it avoids.
Its output is not an admission test but a routing decision.
A failing node is not dropped but handed to the tunneling path, which is what lets \sys check the predicate before any I/O without losing that node as a waypoint.
The check and tunneling therefore depend on each other; checking the filter before I/O without tunneling is exactly the na\"ive joint-filtering that breaks connectivity (\S\ref{sec:bgm}).

The filter store is intentionally decoupled from the graph index.
It is loaded from a separate metadata file and can be replaced independently.
This separation is what enables \sys to support arbitrary predicates (equality, range, multi-label subset, or conjunctions thereof) without rebuilding the graph.
Adding a new predicate type only requires providing the corresponding metadata; the graph index and the search algorithm remain unchanged.
Its memory cost scales with metadata size rather than vector dimensionality (Table~\ref{tab:dram}).

\sys places this gate at the earliest point in the search pipeline, before both hot-node cache lookup and SSD submission.
This placement is safe only because the gate has two exits.
If the check dropped failing nodes instead of routing them, it would remove their routing role along with their result role.
A node that fails therefore incurs neither SSD I/O nor exact distance computation, and bypasses the entire per-read processing chain, not just the disk access.

\noindent\textbf{Neighbor store.}\label{sec:nbr-store}
The \emph{neighbor store} replicates each node's adjacency list from the on-disk graph into memory.
Its purpose is to enable tunneling.
When a node fails the filter, the search needs to access the node's neighbors to continue traversal, but the neighbors are normally embedded in the node's SSD record alongside the full-precision vector.
The neighbor store extracts just the adjacency information (a list of up to $R_{\max}$ neighbor IDs per node) so that tunneling can proceed without any SSD read.

Like the filter store, the neighbor store is built at load time from the existing on-disk index and does not modify it.
Because each node in the Vamana graph stores neighbors in order of proximity, retaining the first $R_{\max}$ entries preserves the closest and most useful neighbors for routing.
The structure supports $O(1)$ lookup by node ID, ensuring that tunneling adds negligible latency compared to the SSD read it replaces.

\begin{figure}[t]
\centering
\includegraphics[width=0.88\linewidth]{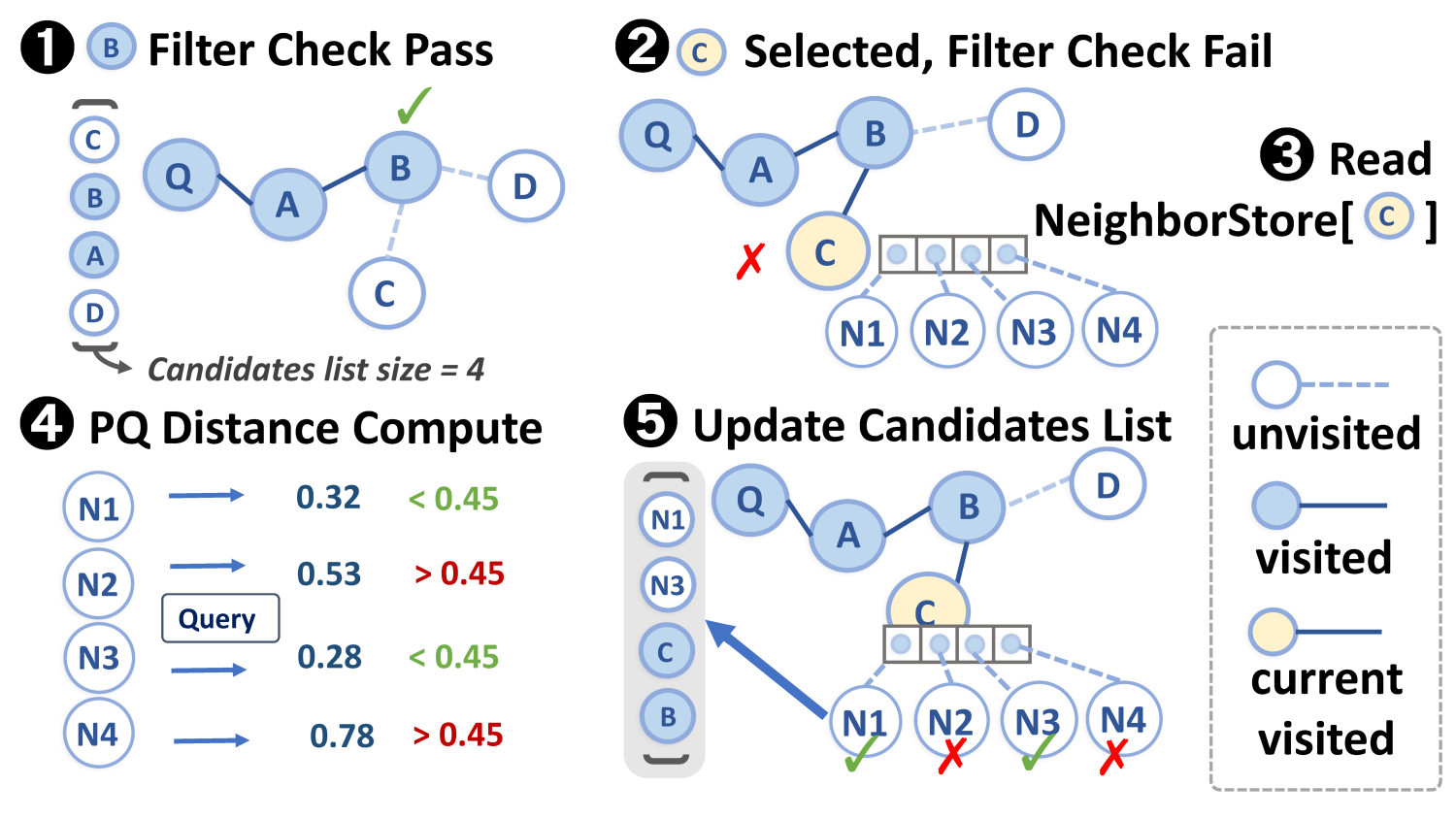}
\caption{Graph tunneling example.
\ding{202}~Node~$B$ passes the filter and follows the normal SSD path.
\ding{203}~Node~$C$ is selected next but fails the filter check.
\ding{204}~\sys reads $C$'s neighbors ($N_1$--$N_4$) from the neighbor store in memory.
\ding{205}~PQ distances to the query are computed for each neighbor; neighbors below the frontier threshold (here $0.45$) are kept ($N_1$, $N_3$) and the rest ($N_2$, $N_4$) are discarded.
\ding{206}~Promising neighbors are inserted into the candidate list, and $C$ is marked visited but ineligible for results.
}
\label{fig:tunneling}
\end{figure}

\subsection{Graph Tunneling}\label{sec:tunneling}
A filter-failing node may still be needed for traversal, so \sys tunnels through it in memory rather than fetching its vector from SSD.
The key observation behind tunneling is that each node in a graph search serves two distinct roles, as a \emph{result candidate} that may appear in the final top-$K$ and as a \emph{routing waypoint} whose edges connect the search to other regions of the graph.
When a filter predicate is present, a non-matching node is ineligible as a result candidate yet may still be essential as a routing waypoint; the two roles diverge.
Post-filtering ignores this divergence, fetching every visited node from SSD regardless of whether it can contribute to the result.
Tunneling separates the two roles.
A node that fails the filter is marked as filter-failing and excluded from the final top-$K$, so it only needs to fulfill its waypoint role.
This role requires only the node's adjacency list and approximate (PQ) distances, both of which reside in memory.

This separation is effective because of a large cost asymmetry between the two paths.
An SSD random read takes on the order of $100\,\mu$s, while a neighbor store lookup and PQ distance computation complete in sub-microsecond time.
Tunneling therefore replaces the most expensive per-node operation with one that is ${\sim}100\times$ cheaper, and the benefit scales with the fraction of visited nodes that fail the filter.

This also clarifies the difference from na\"ive joint-filtering, which skips non-matching nodes without expanding their neighbors and so disconnects the frontier (\S\ref{sec:bgm}).
Tunneling avoids this by preserving neighbor expansion.
It eliminates only the SSD read, not the traversal step, so the search can cross arbitrarily long non-matching regions through a sequence of cheap in-memory hops.

\begin{algorithm}[t]
\caption{\sys search loop (simplified).}\label{alg:search}
\footnotesize
\begin{algorithmic}[1]
\Require Query $q$, filter predicate $f$, search list size $L$, result size $K$, I/O width $W$
\State $\mathcal{C} \gets$ sorted candidate frontier, initialized with the entry point
\State $\mathcal{Q} \gets \emptyset$ \Comment{in-flight SSD reads}
\While{$\mathcal{C}$ has undispatched candidates \textbf{or} $\mathcal{Q} \neq \emptyset$}
  \While{$|\mathcal{Q}| < W$ \textbf{and} $\mathcal{C}$ has an undispatched candidate}
    \State $c \gets$ best undispatched candidate in $\mathcal{C}$
    \If{$\mathit{filter\_store}[c]$ satisfies $f$}
      \State Submit asynchronous read for $c$'s SSD record
      \State Add $c$ to $\mathcal{Q}$ and mark $c$ as dispatched
    \Else
      \State $\mathit{nbrs} \gets \mathit{neighbor\_store}[c]$
      \For{each unvisited $n \in \mathit{nbrs}$}
        \State $d_n \gets \text{PQ\_dist}(q, n)$
        \If{$d_n$ improves the frontier threshold}
          \State Insert $(n, d_n)$ into $\mathcal{C}$
        \EndIf
      \EndFor
      \State Mark $c$ as visited and filter-failing
    \EndIf
  \EndWhile
  \For{each completed read $r$ returned by the SSD}
    \State Compute the exact distance for $r$
    \State Expand $r$'s neighbors from the returned SSD record
    \State Insert promising neighbors into $\mathcal{C}$
    \State Remove $r$ from $\mathcal{Q}$ and mark $r$ as visited
  \EndFor
\EndWhile
\State \Return top-$K$ filter-passing candidates from $\mathcal{C}$
\end{algorithmic}
\end{algorithm}

\noindent\textbf{Example.}
In Figure~\ref{fig:tunneling}, the candidate list is $[C, B, A, D]$.
$B$ passes the filter and takes the SSD path, but $C$ fails.
Instead of reading $C$ from SSD, \sys tunnels through it, scoring $C$'s stored neighbors with PQ distances and inserting the promising ones, which updates the list to $[N_1, N_3, C, B]$.
$C$ never leaves memory, yet still connects the search to new candidates $N_1$ and $N_3$.

\noindent\textbf{Putting it together.}\label{sec:search-algo}
Algorithm~\ref{alg:search} summarizes the integrated search loop.
The important structural property is that both paths, SSD retrieval and in-memory tunneling, feed candidates into the same sorted frontier.
The rest of the search remains unchanged. The algorithm repeatedly selects promising frontier nodes, expands them, and returns the top-$K$ filter-passing candidates.

\subsection{Design Implications and Trade-offs}\label{sec:design-questions}

\noindent\textbf{Approximate distances for tunneling.}\label{sec:tunnel-pq}
Graph tunneling uses PQ distances rather than exact distances from full vectors.
This is acceptable because traversal only needs a good priority signal to decide which neighbors to explore next, not the precise ranking required for the final result set.
Since filter-failing nodes are excluded from the results, the top-$K$ output is always drawn from filter-passing nodes evaluated with exact distances, regardless of the approximation used during traversal.

Two factors may affect tunneling quality.
Tunneled nodes are prioritized with PQ distances, and when $R_{\max} < R$ the in-memory adjacency contains only a subset of each node's neighbors.
This truncation applies only to filter-failing nodes.
A filter-passing node is read from SSD in any case, and the read returns its complete adjacency list of up to $R$ neighbors, so the search expands every passing node at full degree (Algorithm~\ref{alg:search}).
Truncation therefore never removes a node from the graph; it only narrows the out-edges used while routing through a non-matching node.
Neither factor affects the final-result rule above, and PQ errors do not accumulate across hops, since each candidate is rescored independently against the original query and the frontier is globally resorted after every expansion.
When needed, the system can compensate by modestly increasing the search list size $L$, which we evaluate in \S\ref{sec:eval}.

\noindent\textbf{Consecutive non-matching nodes.}\label{sec:tunnel-chain}
At low selectivity, a tunneled node often exposes neighbors that also fail the filter.
\sys handles this case naturally.
If such a neighbor later becomes the best frontier candidate, it is tunneled as well.
In this way, single-hop tunneling composes into multi-hop traversal without any special multi-hop operator.

This design has two advantages.
First, it avoids materializing all nodes within multiple hops of a filtered region, which would grow the work exponentially.
Second, it lets the frontier be re-ranked after every hop, preserving the best-first character of the search.
Long non-matching stretches are a workload-dependent challenge.
If filter predicates are tightly correlated with graph locality, the search may need to cross larger filtered regions before reaching matching nodes, which can require a larger frontier or a larger $R_{\max}$ to maintain recall.
We explicitly investigate these cases in the evaluation rather than assuming label independence.

\noindent\textbf{Memory overhead.}\label{sec:tunnel-dram}
The primary cost is the neighbor store:
\begin{equation}\label{eq:dram}
\text{MEM}_{\text{neighbor}} = N \times (1 + R_{\max}) \times 4~\text{bytes},
\end{equation}
where $N$ is the number of nodes and $R_{\max}$ is the number of in-memory neighbors retained per node.
Table~\ref{tab:dram} shows representative sizes for $N=100$M and $1$B.
Combined with the PQ vectors already kept by the baseline and a modest filter store, the total memory footprint remains practical on commodity servers.
This footprint reduces the largest dataset that a server with fixed memory can hold.
However, the reduction is adjustable at runtime because Eq.~\ref{eq:dram} is linear in $R_{\max}$.
At $R_{\max}{=}8$, the neighbor store costs about as much as the PQ codes the baseline already keeps, and at the default $R_{\max}{=}32$ it costs about four times as much (Table~\ref{tab:dram}).

\sys treats $R_{\max}$ as a \emph{runtime} parameter, not an index-build parameter.
The neighbor store is constructed at load time by scanning the on-disk graph and extracting the first $R_{\max}$ neighbors of each node, without modifying the index itself.
Operators can therefore adjust $R_{\max}$ freely across deployments or restarts without rebuilding the graph.
The load-time scan is a one-time cost per process start, not a per-query cost; it reads the graph file once in order, so it runs at sequential-read bandwidth and grows linearly with the index size, adding one pass to the startup loading \pipeann already performs for PQ vectors and metadata.
Larger values of $R_{\max}$ improve tunneling quality by exposing more candidate routes through non-matching regions, while smaller values reduce memory footprint and per-node in-memory work.
We use $R_{\max}{=}32$ as the default.
At billion scale this keeps the neighbor store at 123\,GB (Table~\ref{tab:dram}), so it fits in a 256\,GB server together with the PQ vectors, whereas $R_{\max}{=}48$ would already require about 183\,GB by Eq.~\ref{eq:dram}; tighter memory budgets can halve the store to 63\,GB with $R_{\max}{=}16$.
Unlike a degree-16 graph that fixes connectivity at build time, \sys truncates only the in-memory routing copy of a full-degree graph, allowing $R_{\max}$ to be increased and reloaded without rebuilding the index.

\begin{table}[t]
\centering
\caption{Memory overhead for $N$-size datasets.}
\label{tab:dram}
\small
\begin{tabular*}{\columnwidth}{@{\extracolsep{\fill}}llrr@{}}
\toprule
Component & Formula & Size (100M) & Size (1B) \\
\midrule
Filter store (single-label) & $N$\,B & 100\,MB & 1\,GB \\
Filter store (multi-label) & varies & $\sim$900\,MB & $\sim$9\,GB \\
Neighbor store ($R_{\max}=32$) & $N \cdot 33 \cdot 4$\,B & 12.3\,GB & 123\,GB \\
Neighbor store ($R_{\max}=16$) & $N \cdot 17 \cdot 4$\,B & 6.3\,GB & 63\,GB \\
Neighbor store ($R_{\max}=8$) & $N \cdot 9 \cdot 4$\,B & 3.4\,GB & 34\,GB \\
\midrule
PQ vectors (baseline) & $N \cdot 32$\,B & 3.2\,GB & 32\,GB \\
\bottomrule
\end{tabular*}
\vspace{-3mm}
\end{table}

\subsection{Discussion}\label{sec:discussion}
\noindent\textbf{Index updates.}
A label change rewrites only the node's filter-store entry, leaving the graph and neighbor store untouched, while a filter-aware index must alter graph edges for the same change.
A vector insertion or deletion rewrites a bounded set of on-disk adjacency lists in dynamic Vamana-style indexes~\cite{odinann}, and since the neighbor store is a truncated copy of these lists, \sys only re-copies the entries the update already rewrote on SSD.
The maintenance cost thus scales with the update's write set rather than the index size, and in the worst case the whole store can be rebuilt by the same sequential scan used at load time.
We leave a concurrent implementation and an evaluation under updates to future work.

\noindent\textbf{Extremely low selectivity.}
Graph traversal is not the best strategy at every selectivity.
When the matching set is small enough, scanning its PQ codes in memory and reranking the top candidates with a few SSD reads costs less than any traversal; prior systems similarly route highly selective queries to specialized structures~\cite{curator,sieve}.
\sys already maintains the two structures this strategy needs, the filter store and the PQ codes, so a per-query router that sends extremely selective queries to an exhaustive scan is a natural extension, which we leave to future work.

%% file: sections/implementation.tex
\begin{table}[t]
\centering
\caption{Datasets used in evaluation.}
\label{tab:datasets}
\footnotesize
\setlength{\tabcolsep}{3pt}
\begin{tabular*}{\linewidth}{@{\extracolsep{\fill}}lrrll@{}}
\toprule
Dataset & Vectors & Dim & Type & Labels \\
\midrule
BigANN-100M & 100M & 128 & uint8 & synth., 10 classes \\
DEEP-100M & 100M & 96 & float & synth., 10 classes \\
YFCC-10M & 10M & 192 & uint8 & real, ${\sim}$200K tags \\
BigANN-1B & 1B & 128 & uint8 & synth., 10 classes \\
\bottomrule
\end{tabular*}
\end{table}

\section{Implementation}\label{sec:impl}
We implement \sys in C++ by extending the \pipeann codebase~\cite{pipeann}.
The implementation adds two read-only metadata structures and modifies the candidate dispatch path in \pipeann's asynchronous search loop.

\noindent\textbf{Filter store.}
The filter store is allocated at index load time from a separate metadata file and is indexed by node ID.
For single-label predicates, each entry stores a fixed-width label.
For multi-label predicates, each entry stores the metadata needed for in-memory predicate evaluation, loaded from a sparse matrix representation.
The structure is read-only during search and shared across all query threads.

\noindent\textbf{Neighbor store.}
The neighbor store stores up to $R_{\max}$ neighbors per node in a contiguous array with fixed stride.
It is built at load time by a sequential scan over the on-disk graph.
We map it with \texttt{mmap} and \texttt{MAP\_POPULATE} so that its pages are materialized before query processing begins.
Without \texttt{MAP\_POPULATE}, the store would page in on demand through the page cache; we materialize it up front so that performance is predictable from the first query.
Like the filter store, the neighbor store is read-only and shared across query threads.

\noindent\textbf{Search loop integration.}
\sys inserts the filter check before \pipeann's SSD submission path.
Candidates that satisfy the predicate follow the original path unchanged.
Candidates that fail the predicate are expanded through the neighbor store using \pipeann's existing PQ distance computation and candidate insertion routines.
The I/O completion path, \texttt{io\_uring} submission and completion logic, and PQ infrastructure are otherwise unchanged.

%% file: sections/evaluation.tex
\begin{figure*}[t]
\centering
\subfloat[BigANN-100M: Recall vs.\ Latency]{%
\includegraphics[width=0.24\linewidth]{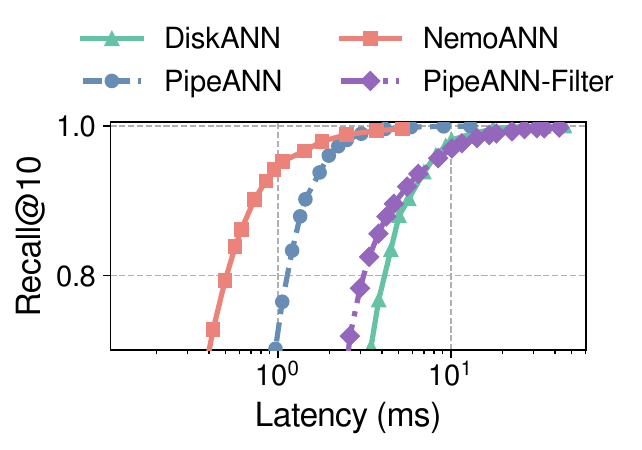}}\hfill
\subfloat[BigANN-100M: QPS vs.\ Recall]{%
\includegraphics[width=0.24\linewidth]{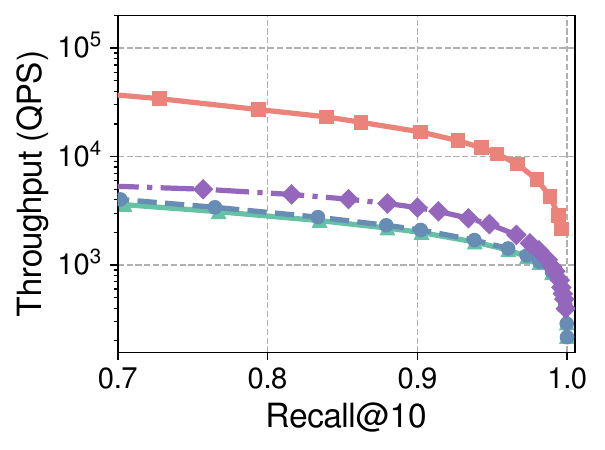}}\hfill
\subfloat[DEEP-100M: Recall vs.\ Latency]{%
\includegraphics[width=0.24\linewidth]{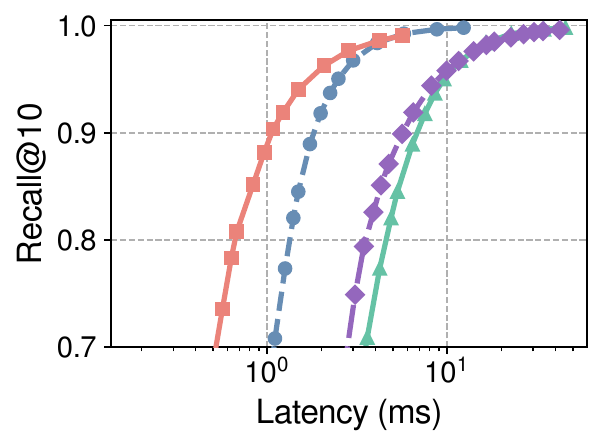}}\hfill
\subfloat[DEEP-100M: QPS vs.\ Recall]{%
\includegraphics[width=0.24\linewidth]{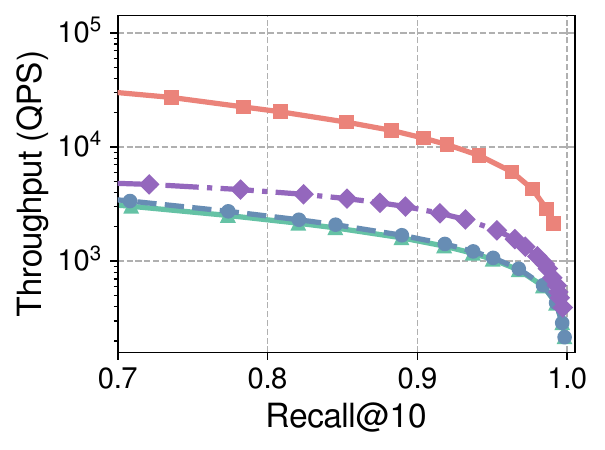}}
\caption{Recall-latency (1 thread) and throughput-recall (32 threads) tradeoff curves on 100M-scale datasets. \sys consistently outperforms both \pipeann and \diskann across the entire recall range.}
\label{fig:pareto-main}
\end{figure*}

\section{Evaluation}\label{sec:eval}

\subsection{Methodology}\label{sec:eval-method}

\noindent\textbf{Testbed.}
All experiments run on a single server with two Intel Xeon Silver 4514Y CPUs (Sapphire Rapids, 32 physical cores supporting 64 logical cores), 256\,GB DDR5, and a Samsung PM9A3 2\,TB NVMe SSD (Gen4).
For \S\ref{sec:eval-ssd}, we additionally use a Samsung 9100 PRO 4\,TB NVMe SSD (Gen5).
The server runs Ubuntu 22.04 with Linux kernel 5.15.

\noindent\textbf{Datasets.}
Table~\ref{tab:datasets} summarizes the used datasets.
BigANN-100M and DEEP-100M use synthetic, uniform 10-class labels, modeling the common category-based filtering scenario.
YFCC-10M~\cite{bigann-neurips23} uses real multi-label metadata ($\sim$200K tag vocabulary) with variable per-query selectivity.
BigANN-1B validates scalability to billion-scale workloads.

\noindent\textbf{Compared systems.}
We compare six systems:
\begin{itemize}[noitemsep,leftmargin=*]
\item \textbf{\diskann}~\cite{diskann}: synchronous beam search ($W{=}8$).
\item \textbf{\pipeann}~\cite{pipeann}: asynchronous pipelined search ($W{=}32$).
\item \textbf{\sys}: asynchronous pipelined search with pre-I/O filter checking and graph tunneling ($W{=}32$, $R_{\max}{=}32$).
\item \textbf{Vamana}~\cite{diskann}: in-memory graph search (no SSD).
\item \textbf{\fdiskann}~\cite{filtered-diskann}: filter-aware graph index with per-label medoids and hard candidate filtering.
\item \textbf{\pipeannfilter}~\cite{pipeann-filter}: speculative filtering on SSD over a densified graph. ($W{=}32$).
\end{itemize}
\diskann, \pipeann, and \sys search the \emph{same} standard Vamana graph index built with graph degree $R{=}96$ and build-time search list $L_{\text{build}}{=}128$, with PQ compression (32 chunks).
\fdiskann uses a \emph{different} index built by FilteredVamana, a label-aware construction that embeds filter information into the graph structure.
\pipeannfilter uses its own densified index built with the code and default settings ($R{=}96$, 2-hop degree $R_d{=}1100$, $L_{\text{build}}{=}128$, 32-byte PQ), so each record occupies 8\,KB sectors.

Note that $W$ has different semantics across systems. In \diskann it is the synchronous beam width (all $W$ reads must complete before the next round), while in \pipeann and \sys it is the asynchronous pipeline depth (up to $W$ concurrent in-flight I/Os).
We follow each system's recommended settings, as in the \pipeann paper~\cite{pipeann}.
\S\ref{sec:eval-bw} shows that \sys's throughput plateaus at $W{\geq}8$, so the specific choice of $W{=}32$ does not inflate our results.

\noindent\textbf{Metrics.}
We report Recall@10, throughput (QPS), mean latency, and mean SSD I/Os per query.
The search list size~$L$ is swept to trace Pareto curves.
Latency experiments use 1 thread to isolate per-query search efficiency without inter-thread contention; throughput experiments use 32 threads to stress the I/O pipeline.
Default selectivity is 10\% unless stated otherwise.

\subsection{Main Results}\label{sec:eval-main}

\subsubsection{Overall Performance}\label{sec:eval-overall}
We show recall-performance curves on BigANN-100M and DEEP-100M in Figure~\ref{fig:pareto-main}.
Each point corresponds to a different search list size~$L$.

\noindent\textbf{Latency.}
On BigANN-100M, \sys achieves ${\sim}$0.73\,ms mean latency at 90\% recall, which is 2.0$\times$ faster than \pipeann, 7.6$\times$ faster than \pipeannfilter, and 7.8$\times$ faster than \diskann.
The improvement is most pronounced at moderate recalls (60--85\%), where the majority of visited candidates fail the filter and are resolved entirely via in-memory tunneling.
At high recall larger than 95\%, \sys remains faster than all baselines, though the gap narrows as the absolute number of filter-passing I/Os grows.
On DEEP-100M, the trend holds with slightly larger absolute latencies due to DEEP's float32 vectors (384 bytes per vector vs.\ BigANN's 128 bytes), which increase the CPU cost of each exact distance computation without affecting \sys's I/O elimination.

\noindent\textbf{Throughput.}
\sys achieves $\sim$16.9K QPS at 90\% recall on BigANN-100M, compared to $\sim$2.1K for \pipeann, an 8.0$\times$ improvement, and $\sim$3.4K for \pipeannfilter, a 5.0$\times$ improvement.
Throughput gains exceed latency gains because 32 threads share a fixed CPU-side I/O processing budget.
Each SSD read triggers not only a disk access but also sector parsing, exact distance computation, and candidate management (\S\ref{sec:eval-breakdown}).
Reduced I/O in \sys frees this budget, enabling throughput to scale with the number of threads rather than per-I/O overhead.
The advantage persists across the recall.
For example, at 95\% recall, \sys sustains a 7.4$\times$ throughput gap over \pipeann and 5.6$\times$ over \pipeannfilter.
\pipeannfilter outperforms \pipeann at 32 threads (its 2-hop neighbors reach recall with fewer reads), but at 1 thread it is slower than \pipeann because each of its reads fetches a two-sector record and expands many 2-hop candidates.

\subsubsection{DRAM-Performance Tradeoff}\label{sec:eval-dram}

\begin{figure}[t]
\centering
\includegraphics[width=0.8\linewidth]{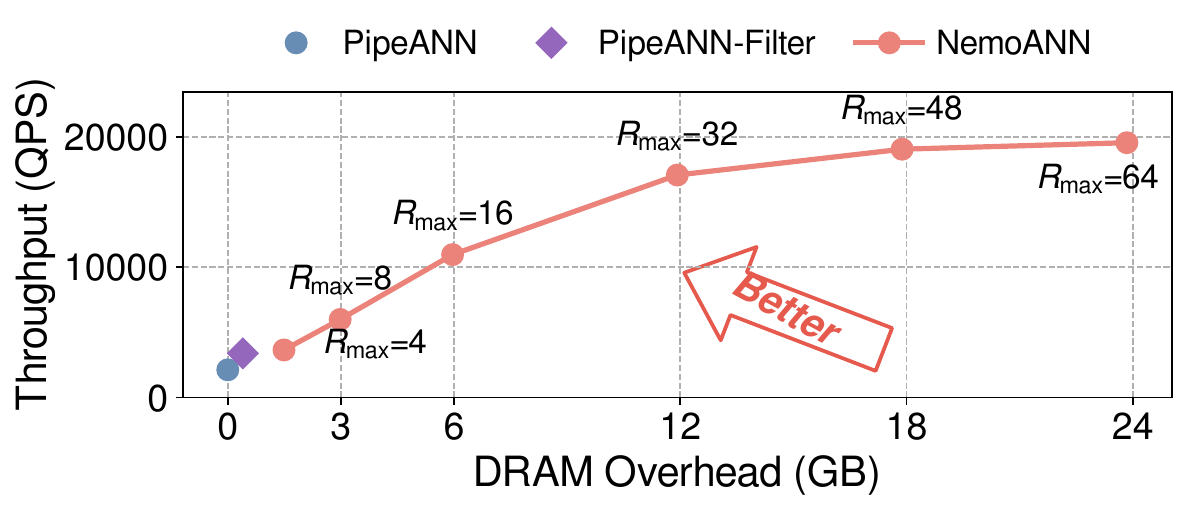}
\caption{Throughput at 90\% recall vs.\ DRAM overhead.}
\label{fig:nbrs-sweep}
\end{figure}
To investigate the trade-off between performance and memory, we sweep $R_{\max}$ from 4 to 64 on BigANN-100M.
Figure~\ref{fig:nbrs-sweep} shows throughput at 90\% recall as a function of DRAM overhead.
Even $R_{\max}{=}8$ at ${\sim}$3.4\,GB delivers 2.8$\times$ higher
throughput than \pipeann.
\pipeannfilter adds only its in-memory label Bloom filter (0.4\,GB) on top of \pipeann and reaches 3.4K QPS; \sys matches it at $R_{\max}{=}4$ (1.9\,GB, 3.7K QPS) and exceeds it by 1.8--5.8$\times$ from $R_{\max}{=}8$ upward.
Throughput saturates beyond $R_{\max}{=}48$ with 19K QPS (9.0$\times$ over \pipeann; 19.5K at 64), while at 1 thread $R_{\max}{=}48$ and 64 are slower than 32 as per-node work grows (\S\ref{sec:tunnel-dram}).
Every configuration in this range outperforms \pipeann at every recall level it reaches, so the memory used by the neighbor store is not a fixed overhead but a tunable knob that lets operators trade memory for performance.
Even its smallest setting adds only a modest footprint over the PQ codes the baseline already keeps.
This sweep also bounds the capacity cost of tunneling.
Memory spent on the neighbor store could otherwise hold a larger dataset, and the knob controls how much; at $R_{\max}{=}8$ the store adds ${\sim}$3.4\,GB, about as much as the PQ codes the baseline itself keeps, yet throughput remains 2.8$\times$ higher than \pipeann.
The alternatives are less attractive, since post-filtering adds no memory but issues an SSD read for every non-matching node, and fully in-memory search uses ${\sim}57$\,GB against \sys's ${\sim}17$\,GB (\S\ref{sec:eval-vamana}).

\begin{figure}[t]
\centering
\subfloat[Mean I/Os per query vs. recall]{%
\includegraphics[width=0.48\linewidth]{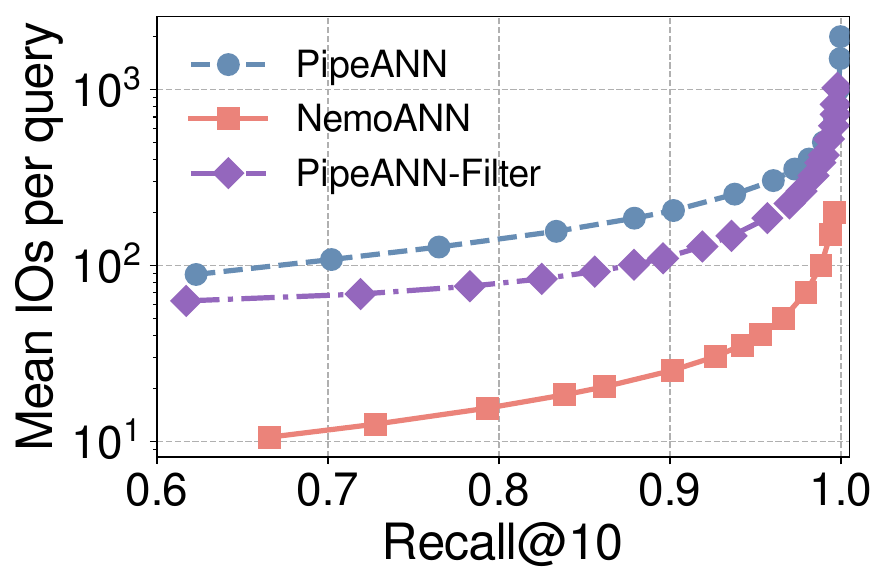}}\hfill
\subfloat[I/O reduction vs.\ selectivity]{%
\includegraphics[width=0.48\linewidth]{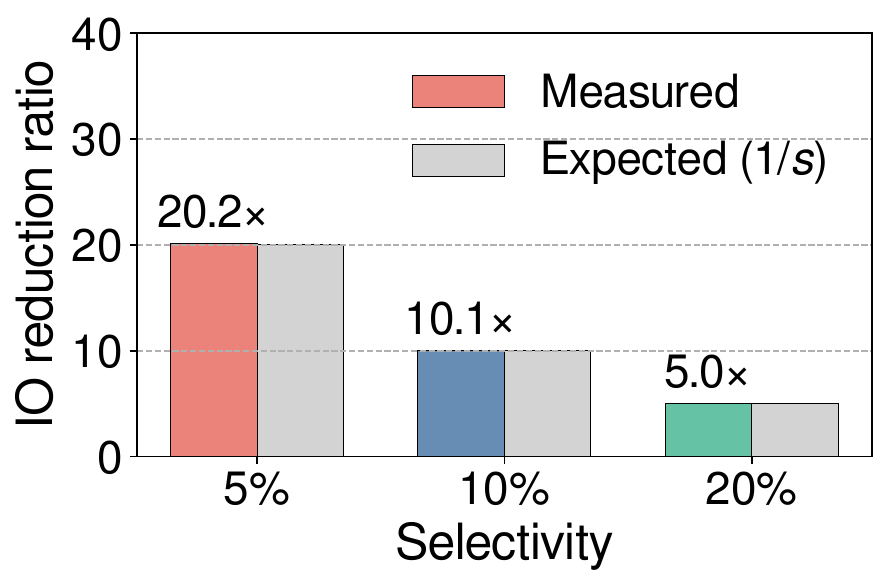}}
\caption{I/O reduction on BigANN-100M. (a)~\sys issues far fewer I/Os than \pipeann and \pipeannfilter at every recall (1 thread; \pipeannfilter reads 8\,KB records, the others 4\,KB). (b)~The measured reduction closely follows the theoretical expectation of $1/s$.}
\label{fig:io-reduction}
\end{figure}

\subsubsection{I/O Reduction}\label{sec:eval-io}
We quantify the SSD I/Os \sys eliminates and whether the reduction matches expectations.
At selectivity~$s$, a fraction~$s$ of visited nodes pass the filter and require SSD I/O; the remaining $1{-}s$ are intercepted by pre-I/O filter checking.
Hence, the expected I/O reduction is $1/s$ compared to a post-filter baseline.

Figure~\ref{fig:io-reduction}~(a) shows the mean I/Os per query as a function of recall on BigANN-100M.
At the same $L$, \sys issues ${\sim}$10$\times$ fewer I/Os than \pipeann (e.g., ${\sim}$206 vs.\ ${\sim}$21 at $L{=}200$); at 90\% recall, \pipeann issues ${\sim}$206 I/Os per query and \pipeannfilter ${\sim}$128, while \sys issues only ${\sim}$25.
The curves stay nearly parallel across the recall range. As $L$ grows, all systems visit proportionally more nodes, but \sys intercepts the same fraction ($1{-}s$) at each step, confirming that the I/O reduction is a structural property of the filter mechanism, not an artifact of a particular~$L$.
Figure~\ref{fig:io-reduction}~(b) compares the measured reduction ratio against the expected $1/s$ at three selectivities: 20.2$\times$ at 5\% (expected: 20$\times$), 10.1$\times$ at 10\% (expected: 10$\times$), and 5.0$\times$ at 20\% (expected: 5$\times$).
The close match validates that pre-I/O filter checking with graph tunneling achieves the optimal I/O count without sacrificing graph connectivity.

\subsubsection{Performance at Billion-Scale}\label{sec:eval-billion}

We evaluate on BigANN-1B.
The 1B index is built with $R{=}128$, $L_{\text{build}}{=}200$ to accommodate the larger graph.
At this scale, the default $R_{\max}{=}32$ requires ${\sim}$123\,GB for the neighbor store; together with PQ codes (${\sim}$32\,GB), the total fits within the server's 256\,GB memory capacity.

\begin{figure}[t]
\centering
\subfloat[Recall vs.\ Latency (1 thread)]{%
\includegraphics[width=0.48\linewidth]{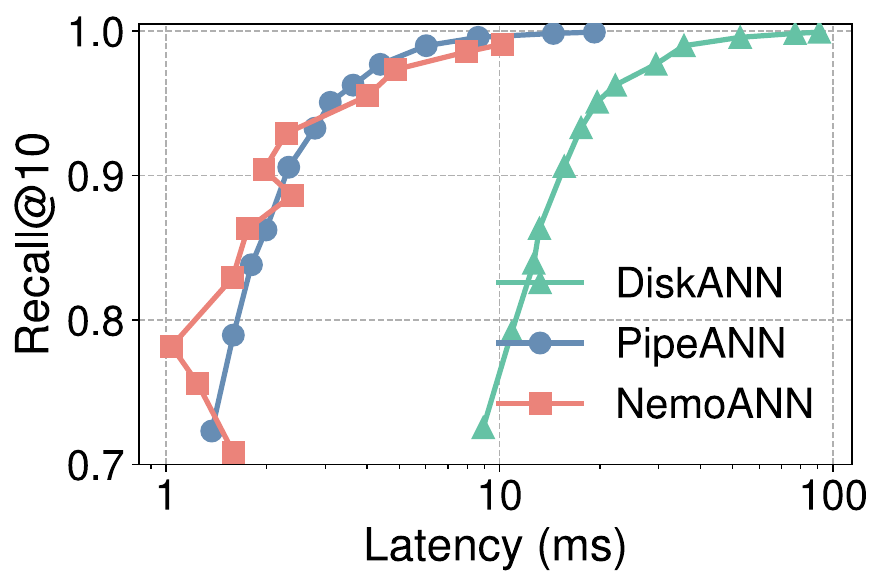}}\hfill
\subfloat[QPS vs.\ Recall (32 threads)]{%
\includegraphics[width=0.48\linewidth]{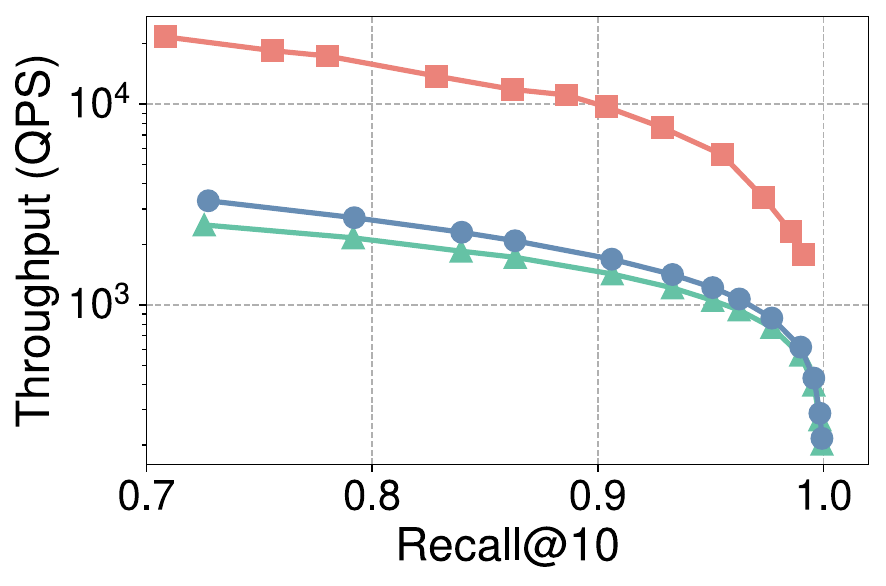}}
\caption{Billion-scale results on BigANN-1B (recall-latency and throughput-recall). \sys's advantage holds at 10$\times$ the scale of the 100M experiments.}
\label{fig:billion}
\end{figure}

Figure~\ref{fig:billion} shows the recall-latency and throughput-recall tradeoff curves.
At 90\% recall with 32 threads, \sys achieves 9.7K QPS, outperforming \pipeann at 1.7K QPS by 5.7$\times$ and \diskann at 1.4K QPS by 6.9$\times$.
This confirms that \sys's pre-I/O filter checking and graph tunneling scale to billion-scale datasets without degradation.
\pipeannfilter is not evaluated at this scale: its densified records occupy two 4\,KB sectors per vector, so a BigANN-1B index would need ${\sim}$8.2\,TB, exceeding the capacity of every SSD in our testbed.

\begin{figure}[t]
\centering
\subfloat[Recall vs.\ Latency (1 thread)]{%
\includegraphics[width=0.48\linewidth]{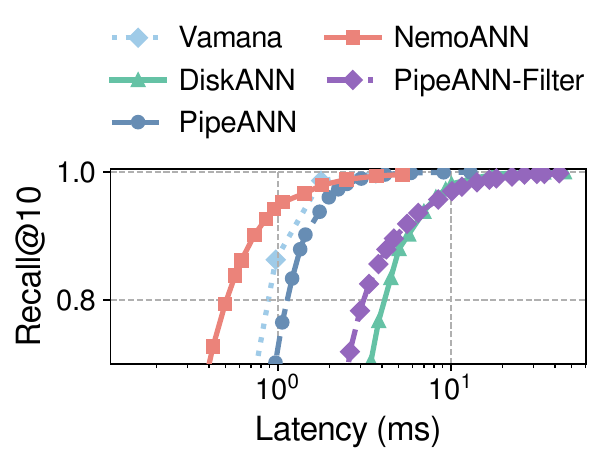}}\hfill
\subfloat[QPS vs.\ Recall (32 threads)]{%
\includegraphics[width=0.48\linewidth]{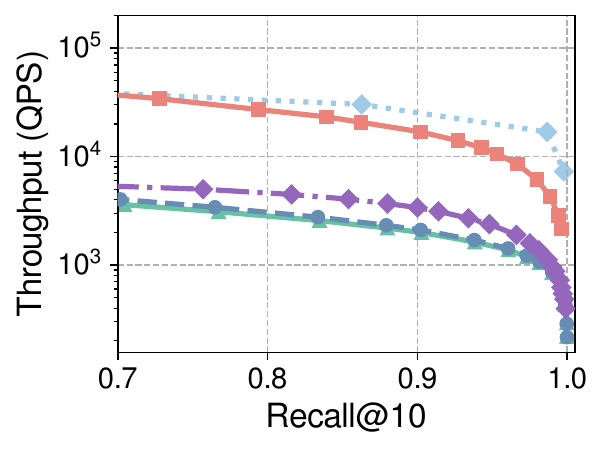}}
\caption{Comparison to Vamana (post-filtering).
}
\label{fig:vamana-analysis}
\end{figure}

\subsubsection{In-Memory Search: Vamana}\label{sec:eval-vamana}
In this experiment, we compare against Vamana~\cite{diskann}, which loads the full graph and all full-precision vectors into memory (${\sim}57$\,GB) and applies post-filtering.
Vamana computes exact distances for every visited node.
Figure~\ref{fig:vamana-analysis}~(a) shows that \sys achieves lower single-thread latency than Vamana at the same recall despite issuing SSD I/Os, because the CPU savings on non-matching nodes outweigh the I/O cost for matching ones.
In Figure~\ref{fig:vamana-analysis}~(b) with 32 threads, Vamana's throughput advantage becomes more apparent as in-memory computation scales freely with more threads while each SSD I/O carries fixed per-operation overhead, yet \sys still reaches within 1.6$\times$ of Vamana at 90\% recall while using only ${\sim}17$\,GB versus ${\sim}57$\,GB, because it keeps full-precision vectors on SSD and needs only compact routing metadata in memory.

\subsubsection{Comparison to Filter-Aware Index}\label{sec:eval-fdiskann}

\begin{figure}[t]
\centering
\subfloat[Recall vs.\ Latency (1 thread)]{%
\includegraphics[width=0.48\linewidth]{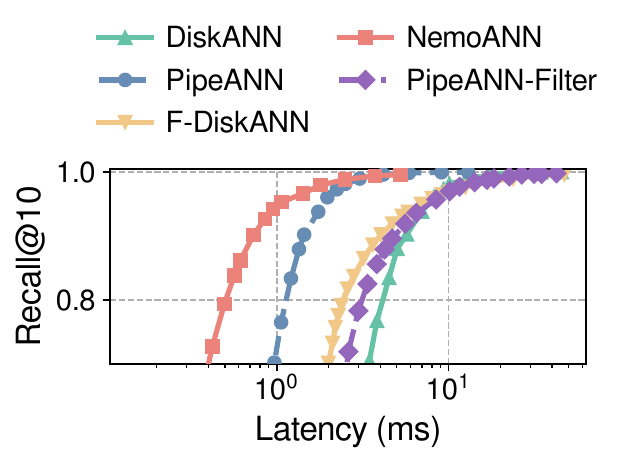}}\hfill
\subfloat[QPS vs.\ Recall (32 threads)]{%
\includegraphics[width=0.48\linewidth]{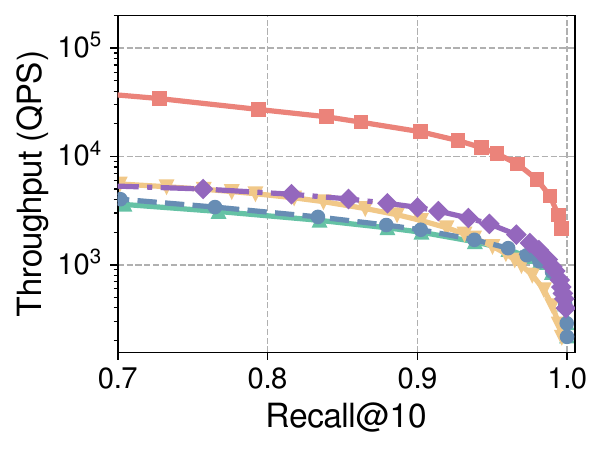}}
\caption{Comparison with \fdiskann and \pipeannfilter. \fdiskann uses a FilteredVamana index with label-aware edge routing and per-label medoid entry points.}
\label{fig:fdiskann}
\end{figure}

\fdiskann~\cite{filtered-diskann} builds a label-aware graph (FilteredVamana) that routes edges preferentially through same-label nodes.
We use the official DiskANN codebase on BigANN-100M with 10-class labels (we only fix a label-file parsing performance bug that otherwise stalls index loading for single-label files).
Figure~\ref{fig:fdiskann} shows that \fdiskann achieves high recall and reduces I/Os by ${\sim}$25\% over \diskann at 90\% recall, confirming a routing benefit.
However, the throughput improvement over \diskann is modest (2.6K vs.\ 2.0K QPS at 32 threads), because although every read serves a matching node, discarding non-matching waypoints lengthens paths between matching nodes, so many more of them must be visited and read.

At 90\% recall, \sys achieves 16.9K QPS, outperforming \fdiskann's 2.6K QPS by 6.6$\times$ and \pipeannfilter's 3.4K QPS by 5.0$\times$ in throughput.
The same trend holds on DEEP-100M, where \sys reaches 12.0K QPS versus \fdiskann's 789 QPS and \pipeannfilter's 2.6K QPS.
Both approaches avoid I/O for non-matching nodes, but \fdiskann pays for it by rebuilding the \emph{graph structure} and fixing the label space at build time, whereas \sys pays with in-memory adjacency in the \emph{search engine} and keeps the original graph and arbitrary predicates.

\subsection{Deep Analysis}\label{sec:eval-deep}

\begin{figure}[t]
\centering
\includegraphics[width=0.9\linewidth]{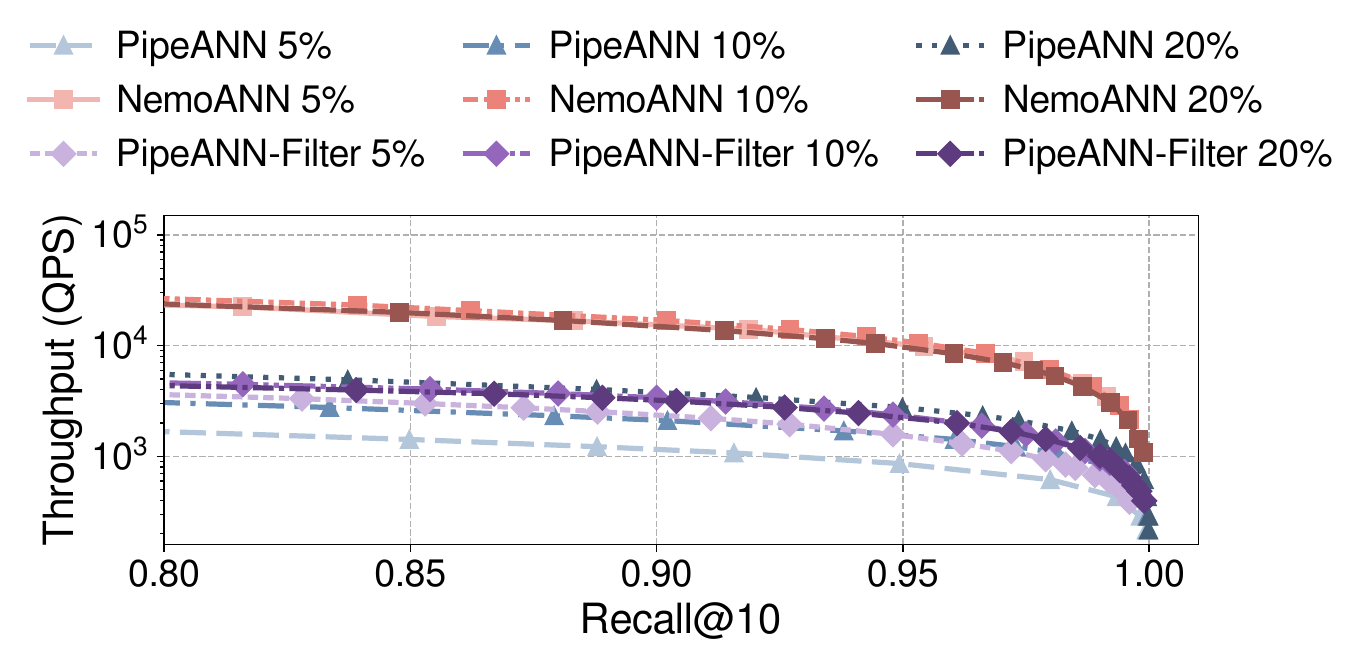}
\caption{QPS vs.\ Recall at 5\%/10\%/20\% selectivity.}
\label{fig:selectivity}
\end{figure}

\subsubsection{Selectivity Sensitivity}\label{sec:eval-sel}
A key property of \sys is that its benefit scales with filter restrictiveness; lower selectivity yields greater I/O reduction.
We evaluate this by varying selectivity across 5\%, 10\%, and 20\% on BigANN-100M with 32 threads.

Figure~\ref{fig:selectivity} shows the throughput--recall tradeoff at each selectivity.
\pipeann's throughput degrades as selectivity decreases, because more nodes must be visited to find enough filter-passing results.
\sys's throughput, in contrast, is largely insensitive to selectivity.
Additional non-matching nodes are resolved via in-memory tunneling, keeping the per-query I/O count low.
At 90\% recall, \sys outperforms \pipeann by 13.1$\times$ at 5\%, 8.0$\times$ at 10\%, and 4.0$\times$ at 20\%, and \pipeannfilter by 6.4$\times$, 5.0$\times$, and 4.3$\times$, respectively.
\pipeannfilter degrades less than \pipeann as selectivity drops, but still reads 4--8$\times$ more records than \sys at 90\% recall.
The I/O reduction tracks $1/s$ (\S\ref{sec:eval-io}), while the throughput gain grows more slowly because tunneling adds CPU work for every non-matching candidate.
Two further experiments probe below the 5\% point.
The Zipf workload (\S\ref{sec:eval-zipf}) includes rare-class queries at 3.4\% selectivity, and most YFCC queries (\S\ref{sec:eval-yfcc}) require rare tags, which pulls the median per-query selectivity to 0.15\%.
In both cases \sys retains its advantage over \pipeann.
On YFCC it also reaches a higher maximum recall (91.5\%) than \pipeann (83\%), although \pipeann expands every visited node with its full on-disk adjacency; the truncated in-memory lists therefore do not cap recall below what the full-adjacency baseline reaches.

\begin{figure}[t]
\centering
\includegraphics[width=0.82\linewidth]{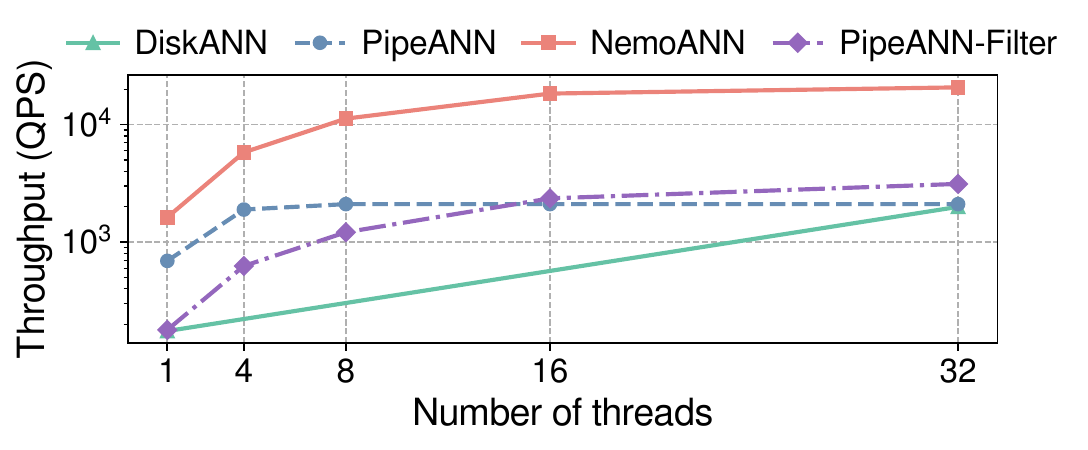}
\caption{Throughput scaling from 1 to 32 threads on BigANN-100M, $L{=}200$ (\pipeannfilter at the $L$ that matches \pipeann's recall).}
\label{fig:thread-scaling}
\end{figure}

\subsubsection{Scalability}\label{sec:eval-thread}
Figure~\ref{fig:thread-scaling} reports throughput for 1--32 threads at $L{=}200$.
\pipeann plateaus at ${\sim}$2.1K QPS from 8 threads, and \diskann converges to the same level (${\sim}$2.0K) at 32 threads despite \pipeann's 4.0$\times$ single-thread advantage, because both saturate the same aggregate I/O processing budget of ${\sim}$430K IOPS.
\pipeannfilter (plotted at the $L$ that matches \pipeann's recall) scales from 0.18K to 3.1K QPS and gains little beyond 16 threads.
\sys does not raise this ceiling but needs far less of it: it issues ${\sim}$21 instead of ${\sim}$206 I/Os per query, and within the saturated budget throughput is inversely proportional to I/Os per query, so its gap over \pipeann widens from 2.3$\times$ at 1 thread to 9.9$\times$ at 32 threads (20.7K QPS), matching the 10$\times$ I/O reduction.
\sys itself scales sublinearly (12.8$\times$) for the same reason, as its aggregate rate at 32 threads approaches the same ${\sim}$430K IOPS budget.
Because each thread issues its next query as soon as the previous one completes, these curves also determine mean latency; once \pipeann plateaus, each additional thread only lengthens its latency.

\subsubsection{Impact of SSD Speed}\label{sec:eval-ssd}

A natural question is whether faster SSDs diminish the benefit of \sys.
We compare the Gen4 PM9A3 and Gen5 Samsung 9100 PRO, which differ by approximately $2\times$ in random 4\,KB read throughput.

Table~\ref{tab:ssd-impact} reveals each SSD sensitivity.
At 1~thread, \diskann benefits the most (1.53$\times$) because every query serially waits for I/O; \pipeann's asynchronous pipeline already hides most device latency, limiting its gain to 1.10$\times$; and \sys sees no benefit (0.99$\times$) because its ${\sim}$20 I/Os per query leave the bottleneck entirely at CPU.
At 32 threads, \pipeann shows \emph{zero} benefit from the faster SSD.
This is the key result.
Were the SSD the saturated resource, doubling its IOPS capacity would improve throughput proportionally, so the zero gain shows that the binding constraint is on the host side; Table~\ref{tab:breakdown} attributes it to the processing chain triggered by each read.
For reference, the Gen4 device is rated at over 800K random 4\,KB read IOPS, roughly twice the ${\sim}$430K IOPS ceiling, so the device itself is never the saturated resource in our configurations.
\diskann benefits modestly (1.06$\times$) because its synchronous batching has lower per-I/O CPU overhead, leaving device latency a nontrivial share of its time, while \sys remains SSD-independent (1.04$\times$) simply because it has already eliminated the wasted reads.
Faster SSDs alone thus cannot solve the filtered search problem.
The effective lever is reducing the \emph{number} of I/Os, not the \emph{speed} of each.

\begin{table}[t]
\centering
\caption{Performance on Gen4 vs.\ Gen5 SSDs (BigANN-100M, Recall@10 $\geq$ 90\%). The Gen5/Gen4 column shows how much each system benefits from the faster SSD.}
\label{tab:ssd-impact}
\small
\begin{tabular}{lcc}
\toprule
 & Gen5/Gen4 (1 thread) & Gen5/Gen4 (32 threads) \\
\midrule
\diskann  & 1.53$\times$ & 1.06$\times$ \\
\pipeann  & 1.10$\times$ & 1.00$\times$ \\
\sys      & 0.99$\times$ & 1.04$\times$ \\
\bottomrule
\end{tabular}
\end{table}

\subsubsection{Time Breakdown}\label{sec:eval-breakdown}

\begin{table}[t]
\centering
\caption{Per-query time breakdown at ${\sim}$86--90\% recall (1 thread, BigANN-100M).}
\label{tab:breakdown}
\footnotesize
\setlength{\tabcolsep}{3pt}
\begin{tabular}{@{}lrrrr@{}}
\toprule
Component & \pipeann ($\mu$s) & \% & \sys ($\mu$s) & \% \\
\midrule
SSD I/O (submit + poll)    & 75  & 5.2  & 10  & 1.6 \\
Tunneling (PQ + AdjIndex) & --- & ---  & 311 & 50.5 \\
Processing (exact dist.)  & 991 & 68.6 & 109 & 17.7 \\
Other (list mgmt., loop)  & 379 & 26.2 & 185 & 30.1 \\
\midrule
Total                     & 1445 & 100 & 615 & 100 \\
\bottomrule
\end{tabular}
\setlength{\tabcolsep}{6pt}
\end{table}

Table~\ref{tab:breakdown} decomposes per-query latency at ${\sim}$86--90\% recall (1 thread).
In \pipeann, exact distance computation dominates (68.6\%), while SSD I/O itself is only 5.2\%.
Each I/O triggers an expensive processing chain that far exceeds the device access time.
\sys replaces this path with in-memory tunneling for non-matching nodes, shrinking processing by 9.1$\times$ from 991 to 109\,$\mu$s and total latency by 2.3$\times$ from 1445 to 615\,$\mu$s.

At 32~threads, the gap widens further.
Because each I/O carries the full processing chain above (submission, polling, sector parsing, and distance computation), 32 threads collectively exhaust their CPU budget at ${\sim}$430K aggregate IOPS, a CPU-side ceiling rather than an SSD limit (\S\ref{sec:eval-ssd}).
Beyond this ceiling, throughput becomes inversely proportional to I/Os per query, which explains the 9.9$\times$ throughput gap from the 10$\times$ I/O reduction reported in \S\ref{sec:eval-thread}.

\subsubsection{An Extreme Case: Multi-Label YFCC-10M}\label{sec:eval-yfcc}

\begin{figure}[t]
\centering
\includegraphics[width=0.74\linewidth]{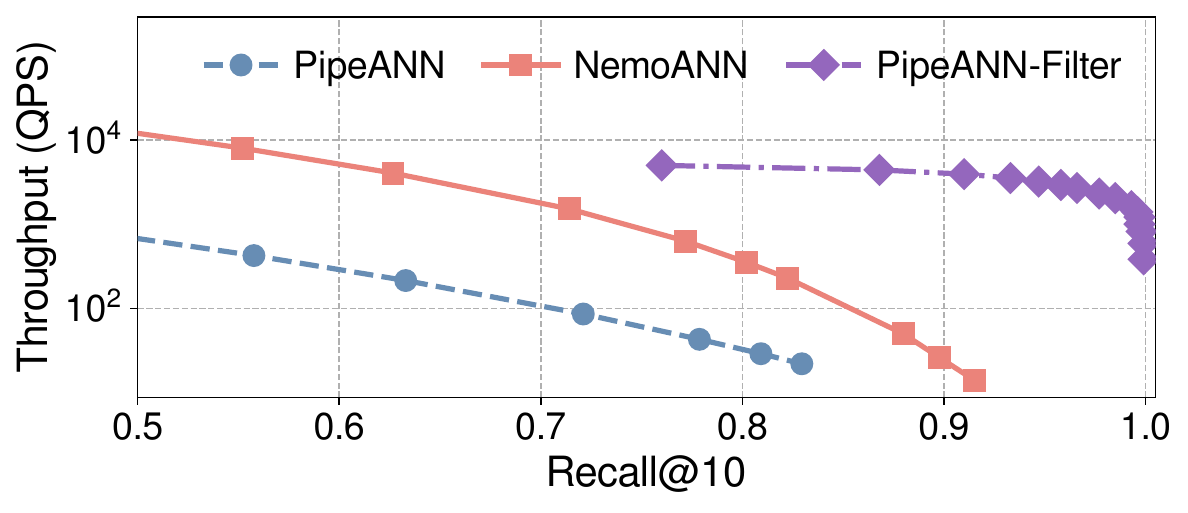}
\caption{Throughput--recall tradeoff curves on YFCC-10M, an extreme multi-label workload (conjunctive tag predicates; median per-query selectivity 0.15\%).}
\label{fig:yfcc}
\end{figure}

We further evaluate \sys on an extreme workload, YFCC-10M from the BigANN benchmark~\cite{bigann-neurips23}, where results must contain all query tags from a $\sim$200K-term vocabulary.
Such conjunctions make most queries extremely selective, and the median query matches only 0.15\% of vectors (mean 1.9\%).
Against post-filtering, \sys's advantage grows further.
It achieves 16.9$\times$ higher throughput than \pipeann at ${\sim}$35\% recall with 18.7$\times$ fewer I/Os, and reaches 91.5\% recall, whereas \pipeann stalls at 83\% even at $L{=}$20K.

\pipeannfilter, however, is far faster here, reaching 91\% recall at 4.0K QPS.
Its cost model routes 42--75\% of the queries to pre-filtering, which ranks the matching set from its label index with in-memory PQ codes and bypasses graph traversal entirely.
When a query matches only hundreds to thousands of vectors, such a scan is cheaper than any traversal.
This advantage reflects an extreme setting.
Pre-filtering cost grows with the matching-set size $s \cdot N$, so the small 10M scale keeps the median matching set at ${\sim}$15K vectors, whereas its authors report pre-filtering falling below 100 QPS once a query matches over a million vectors~\cite{pipeann-filter}.
\sys instead focuses solely on graph traversal and targets the 1--10\% selectivities common in practice (\S\ref{sec:io-waste}), where matching sets reach millions of vectors already at the 100M scale.
A pre-filtering path for extremely selective queries is therefore complementary to \sys.

\subsubsection{Skewed Label Distributions}\label{sec:eval-zipf}

Real-world label distributions are typically skewed; a few labels are common while many are rare.
We test robustness by replacing uniform labels with a Zipf distribution ($\alpha{=}1.0$, 10 classes), where the most common class covers 34\% of vectors and the rarest covers only 3.4\%.
Queries are distributed uniformly across classes, creating a mix of high- and low-selectivity queries in each run.

\begin{figure}[t]
\centering
\subfloat[Recall vs.\ Latency (1 thread)]{%
\includegraphics[width=0.48\linewidth]{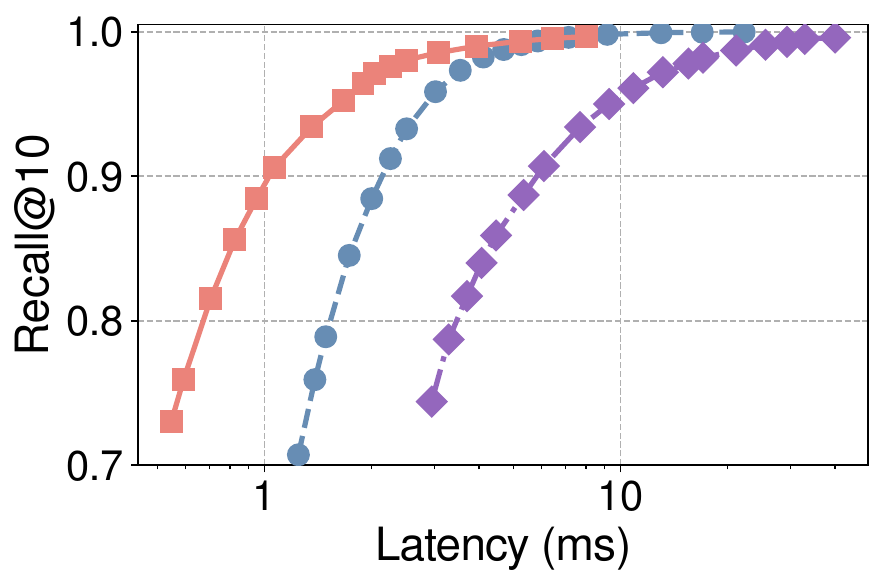}}\hfill
\subfloat[QPS vs.\ Recall (32 threads)]{%
\includegraphics[width=0.48\linewidth]{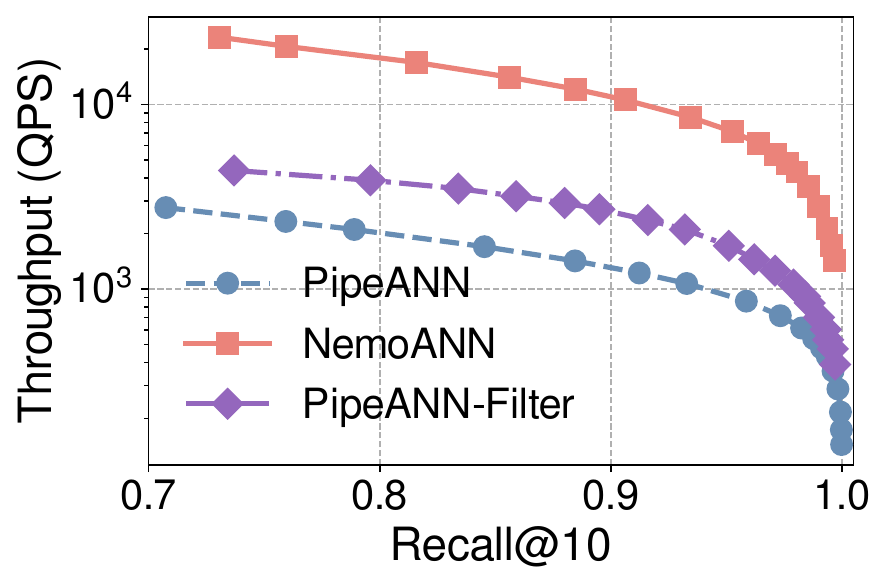}}
\caption{BigANN-100M with Zipf-distributed labels ($\alpha{=}1.0$, 10 classes). Selectivity ranges from 3.4\% (rare) to 34.1\% (common) across queries.}
\label{fig:zipf}
\end{figure}

Figure~\ref{fig:zipf} shows that \sys maintains its advantage under skewed distributions.
The mixed selectivities are favorable for \sys.
Rare-class queries (3.4\% selectivity) see even larger I/O reductions, while common-class queries (34\%) still benefit because two-thirds of candidates fail the filter.
The overall throughput improvement of 8.7$\times$ over \pipeann and 4.5$\times$ over \pipeannfilter at 90\% recall reflects the selectivity-weighted average.

\subsubsection{Spatial Label Correlation}\label{sec:eval-spatial}

\begin{figure}[t]
\centering
\includegraphics[width=0.9\linewidth]{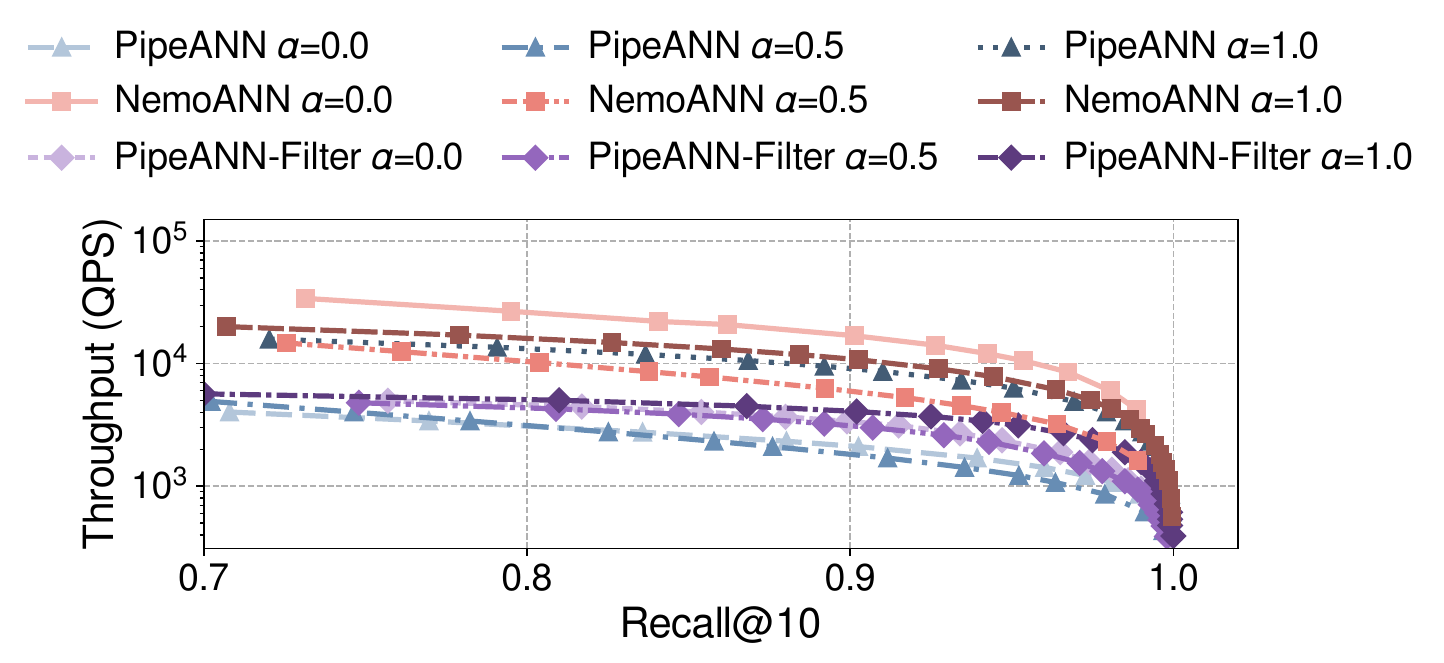}
\caption{Effect of label-vector correlation. Labels are assigned via $k$-means; $\alpha{=}0$ is random, and $\alpha{=}1$ assigns each node the label of its nearest cluster center.}
\label{fig:spatial}
\end{figure}

The preceding experiments assign labels uniformly at random, but real-world metadata often correlates with the vector space.
To measure the impact of such correlation, we assign labels using $k$-means clustering ($k{=}10$ on BigANN-100M) with a mixing parameter~$\alpha$.
At $\alpha{=}0$ labels are random; at $\alpha{=}1$ each node receives the label of its nearest cluster center.
Selectivity remains 10\% (10 classes) at all~$\alpha$.

Figure~\ref{fig:spatial} shows the results.
Recall in this experiment and in \S\ref{sec:eval-range} is measured against the exact filtered ground truth (\S\ref{sec:eval-method}), so no reference-induced recall ceiling arises.
All systems reach at least 98.7\% recall at every~$\alpha$, even though \sys routes through non-matching nodes using only its truncated in-memory lists.
At 90\% recall, \sys outperforms \pipeann at every~$\alpha$ (8.0$\times$ at $\alpha{=}0$, 3.1$\times$ at 0.5, 1.2$\times$ at 1), but the gap shrinks as correlation increases.
Against \pipeannfilter, the gap shrinks from 5.4$\times$ at $\alpha{=}0$ to 1.8$\times$ at $\alpha{=}0.5$ and near parity at $\alpha{=}0.75$ (not plotted), before widening again to 2.6$\times$ at $\alpha{=}1$.
At $\alpha{=}0$, \sys eliminates ${\sim}$90\% of I/Os; at $\alpha{=}1$, traversal naturally stays within the matching cluster, leaving fewer wasted I/Os to eliminate.

\subsubsection{Range Predicates}\label{sec:eval-range}

\begin{figure}[t]
\centering
\subfloat[Recall vs.\ Latency (1 thread)]{%
\includegraphics[width=0.48\linewidth]{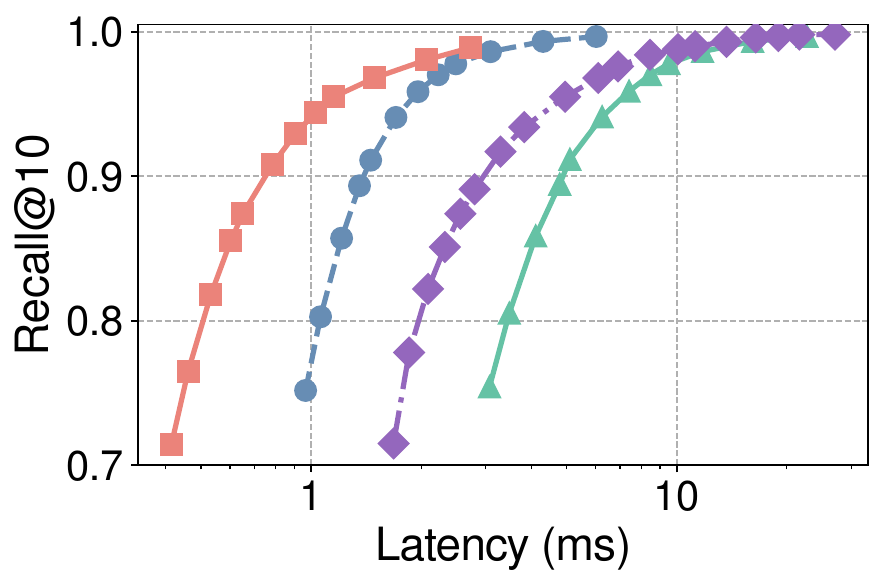}}\hfill
\subfloat[QPS vs.\ Recall (32 threads)]{%
\includegraphics[width=0.48\linewidth]{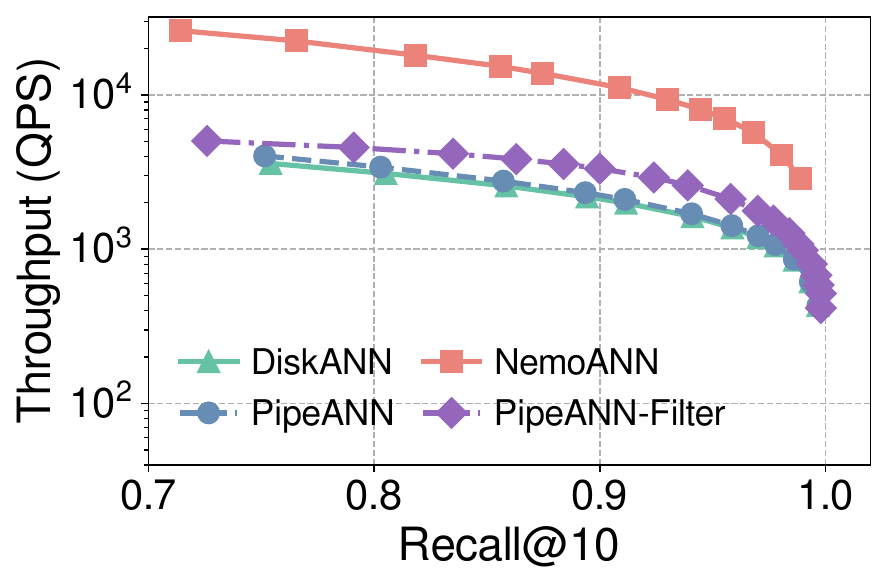}}
\caption{Range predicate (L2-norm binning, 10 bins).}
\vspace{-1pt}
\label{fig:range}
\end{figure}

To validate that \sys handles non-categorical predicates, we construct a range predicate by binning each vector's L2 norm into 10 equal-frequency bins and selecting one bin per query (${\sim}$10\% selectivity).
This predicate is geometric and creates concentric-shell-like regions in the vector space.

Figure~\ref{fig:range} shows latency and throughput curves.
At 90\% recall with 32 threads, \sys achieves 11.2K QPS, 5.3$\times$ higher than \pipeann's 2.1K QPS and 3.4$\times$ higher than \pipeannfilter's 3.3K QPS; with 1 thread, \sys reaches 90\% recall at ${\sim}$0.78\,ms vs.\ ${\sim}$1.5\,ms for \pipeann, ${\sim}$3.3\,ms for \pipeannfilter, and ${\sim}$5.1\,ms for \diskann.
Measured against the exact ground truth, all systems approach full recall (\sys reaches 98.9\%).
The result confirms that \sys is predicate-agnostic.
It applies equally to range predicates without any change to the index or search algorithm.

\begin{figure}[t]
\centering
\subfloat[QPS vs.\ $W$ at 93\% recall]{%
\includegraphics[width=0.48\linewidth]{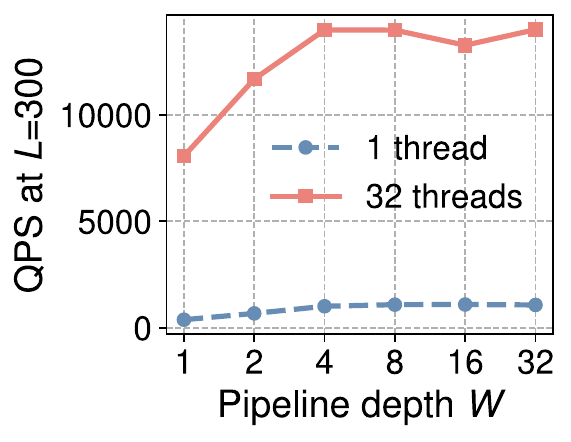}}\hfill
\subfloat[Recall is invariant to $W$]{%
\includegraphics[width=0.48\linewidth]{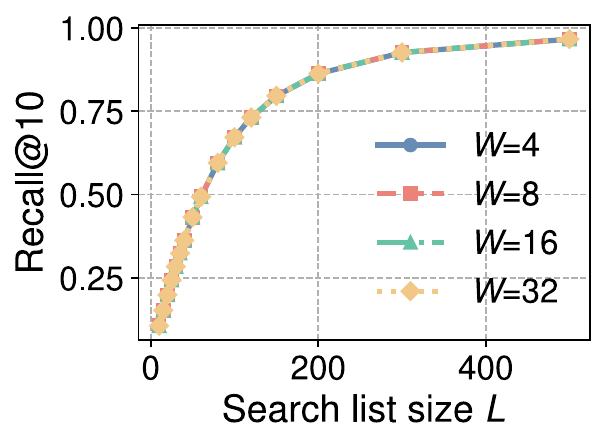}}
\caption{Pipeline depth ($W$) sweep for \sys on BigANN-100M. Recall is identical across all~$W$; throughput plateaus at $W{\geq}8$ (32 threads) and at $W{\geq}4$ (1 thread).}
\label{fig:bw-sweep}
\end{figure}
\subsubsection{Pipeline Depth}\label{sec:eval-bw}
In \pipeann, pipeline depth~$W$ controls the number of in-flight I/Os via \texttt{io\_uring}.
Since our solution issues ${\sim}10{\times}$ fewer I/Os per query, one might expect a smaller~$W$ to suffice.
We sweep $W \in \{1, 2, 4, 8, 16, 32\}$ on BigANN-100M.

Figure~\ref{fig:bw-sweep}~(b) confirms that recall is invariant to~$W$. All values produce identical curves across~$L$, since $W$ affects only I/O scheduling, not candidate selection.
Figure~\ref{fig:bw-sweep}~(a) shows throughput at $L{=}300$ (${\sim}$93\% recall).
At 32 threads, throughput improves from $W{=}1$ (8.8K QPS) to $W{=}4$ (14.1K QPS), then plateaus through $W{=}32$ (14.0K QPS).
The remaining I/Os per query still benefit from concurrency, but 4 concurrent I/Os suffice to hide per-read latency, even though \sys issues ${\sim}$10$\times$ fewer of them.
At 1~thread, throughput plateaus at $W{\geq}8$ (${\sim}$1.1K QPS), confirming that a single thread is CPU-bound even at moderate depths.

\subsubsection{Ablation: I/O Elimination vs.\ CPU Savings}\label{sec:eval-ablation}

\begin{figure}[t]
\centering
\subfloat[Recall vs.\ Latency (1 thread)]{%
\includegraphics[width=0.48\linewidth]{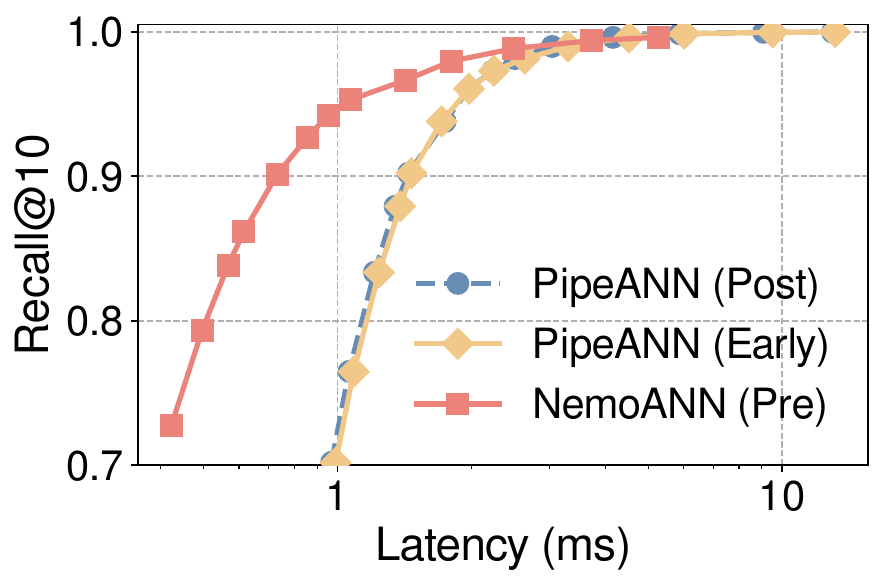}}\hfill
\subfloat[QPS vs.\ Recall (32 threads)]{%
\includegraphics[width=0.48\linewidth]{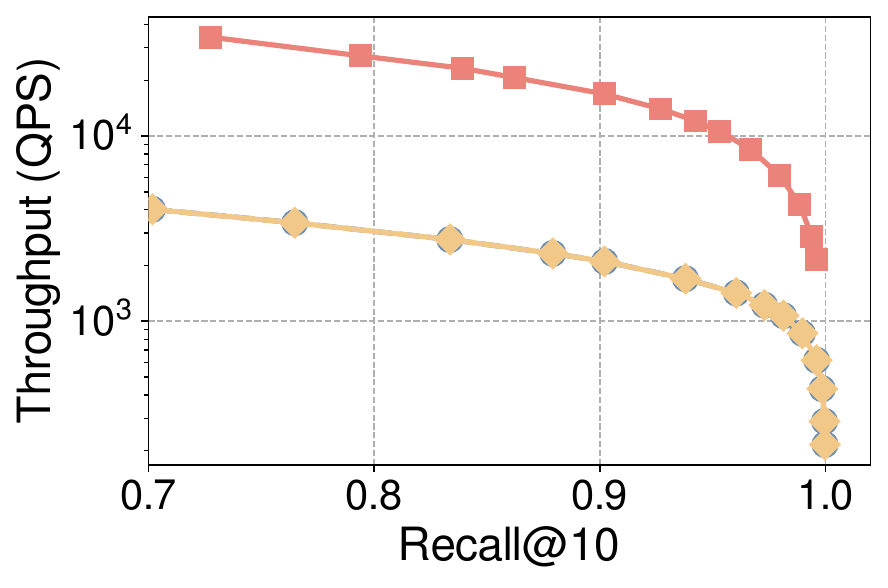}}
\caption{Ablation on BigANN-100M ($s{=}0.1$). \emph{PipeANN~(Early)} filters after each SSD read, skipping exact distance computation for non-matching nodes; \emph{\sys~(Pre)} additionally eliminates the SSD reads themselves.}
\label{fig:early-filter}
\end{figure}

\sys's performance gains come from two sources: (i) eliminating SSD I/Os for non-matching nodes, and (ii) skipping exact distance computation on them.
To separate them, we add PipeANN~(Early), which filters \emph{after} the read.
Non-matching nodes still incur the I/O but skip the exact distance, while neighbor expansion is unchanged to preserve connectivity.

Figure~\ref{fig:early-filter} compares the three designs.
At 1 thread (Figure~\ref{fig:early-filter}(a)), PipeANN~(Early) and PipeANN~(Post) are nearly identical.
At 90\% recall, latency is 1.47\,ms vs.\ 1.45\,ms, within measurement noise.
Skipping exact distance saves only tens of $\mu$s per query because BigANN's 128-D vectors make each of ${\sim}$206 computations cheap (${\sim}$90\,ns).
In contrast, \sys~(Pre) (i.e., \sys~(Joint)) reduces latency to 0.73\,ms (2.0$\times$) and reaches $\sim$16.9K QPS (8.0$\times$) by eliminating SSD reads.
Removing distance computation without removing I/O does not address the bottleneck.
As Table~\ref{tab:breakdown} shows, per-I/O cost is dominated by submission, polling, and sector parsing, and both PipeANN variants still issue ${\sim}$206 I/Os per query, hitting the same ${\sim}$430K IOPS ceiling.
Only \sys eliminates the SSD reads themselves, confirming that \emph{what to read} matters far more than \emph{what to compute}.

%% file: sections/related.tex
\section{Related Work}\label{sec:related}

\noindent\textbf{SSD-based graph ANNS.}
DiskANN~\cite{diskann} introduced SSD-resident graph search with batched beam search.
Subsequent work improves SSD efficiency through asynchronous I/O pipelines~\cite{pipeann}, locality-aware layout~\cite{starling,margo}, layout-runtime co-design~\cite{veloann}, page-aligned placement~\cite{pageann}, and direct-insert updates~\cite{odinann}.
GPS~\cite{gps} decouples adjacency from exact vectors by caching full adjacency for offline-selected nodes, but targets unfiltered, query-independent layout optimization through disk-layout changes.
Neither the inverted-index SPANN~\cite{spann} nor the cluster-partitioned OrchANN~\cite{orchann} integrates predicate-aware filtering, so metadata constraints fall back to post-filtering.
\sys instead avoids fetching non-matching nodes and is complementary to any on-disk graph index.

\noindent\textbf{Filtered ANNS.}
Most work is in-memory, including approaches that embed attributes into the index by fusing vector-attribute distances~\cite{nhq, jag}, by stitching per-label-set subgraphs via label containment~\cite{ung}, or by densifying edges to keep predicate-induced subgraphs navigable~\cite{acorn}.
Others maintain multiple filter-specific subindexes selected at query time~\cite{curator, sieve}.
For disk-based systems, Filtered-DiskANN~\cite{filtered-diskann} builds a label-aware Vamana graph with per-label medoids.
NaviX~\cite{navix} uses predicate-agnostic pre-filtering on disk-based HNSW index, but evaluates the predicate over the full dataset per query and keeps adjacency on disk.
PipeANN-Filter~\cite{pipeann-filter} is a recent work that explores a superset of valid nodes and verifies at rerank time, but storing 10--20$\times$ extra 2-hop neighbors inflates its on-disk footprint, making billion-scale serving impractical.
\sys keeps the original graph and serves non-matching nodes as in-memory routing waypoints, eliminating their SSD reads without rebuilding the index or its on-disk layout.

%% file: sections/conclusion.tex
\section{Conclusion}\label{sec:conclusion}
This paper presented \sys, an I/O-efficient SSD-based system for filtered vector search.
A node that fails the filter contributes nothing to the results, so the search consults only its neighbor list.
That list is small enough to keep in memory.
\sys exploits this by evaluating the filter before every SSD read; a failing node is traversed through an in-memory copy of its neighbor list, so its read is avoided and graph connectivity is preserved.
In our experiments, \sys reduces I/Os by about 10$\times$ and improves throughput by 8.0$\times$ on a 100M-vector benchmark at 10\% selectivity and 90\% recall, with larger gains at lower selectivity.

\begin{acks}
The authors would like to thank the anonymous reviewers for their
insightful comments and feedback. This research was sponsored by the
National Research Foundation of Korea (NRF) grants funded by the
Ministry of Science and ICT (No. RS-2025-00522990).
Gyuyeong Kim is the corresponding author.
\end{acks}